\documentclass[usenatbib]{mn2e}
\usepackage{graphicx}
\usepackage{bm}
\usepackage{fixltx2e}
\usepackage{astrobib_mnras2e}

\begin{document}
\topmargin-1cm

\newcommand{\bd}{{\bm \delta}}

\title[A 2\% Distance to $z=0.35$ : Methods and Data]
{A 2\% Distance to $z=0.35$ by Reconstructing Baryon Acoustic Oscillations - I : Methods and 
Application to the Sloan Digital Sky Survey}
\author[Padmanabhan et al]{Nikhil Padmanabhan$^{1}$, Xiaoying Xu$^{2}$, 
Daniel J. Eisenstein$^{3}$,
Richard Scalzo$^{1,4}$, \newauthor
Antonio J. Cuesta$^{1}$, 
Kushal T. Mehta$^{2}$, 
Eyal Kazin$^{5, 6}$ \\
$^{1}$ Dept. of Physics, Yale University, 260 Whitney Ave, New Haven, CT 06520 \\ 
$^{2}$ Steward Observatory, University of Arizona, 933 N. Cherry Ave., Tucson, AZ 85721 \\
$^{3}$ Harvard-Smithsonian Center for Astrophysics, Harvard University, 60 Garden St., Cambridge, MA 02138 \\
$^{4}$  Research School of Astronomy \& Astrophysics,The Australian National University,
  Mount Stromlo Observatory, \\Cotter Road, Weston ACT 2611, Australia \\
$^{5}$ Centre for Astrophysics and Supercomputing, Swinburne University of Technology,  
     P.O. Box 218,   Hawthorn, Victoria 3122, Australia \\
$^{6}$ Center for Cosmology and Particle Physics, New York University, 4 Washington Place, 10003 NY, USA \\
}

\date{\today}
\maketitle

\begin{abstract}
We apply the reconstruction technique to the clustering of galaxies
from the Sloan Digital Sky Survey Data Release 7 (SDSS DR7) Luminous
Red Galaxy sample, sharpening the baryon acoustic oscillation (BAO) feature
and achieving a 1.9\% measurement of the distance to $z=0.35$.  This
is the first application of reconstruction of the BAO feature in a
galaxy redshift survey.
We update the reconstruction algorithm of \cite{2007ApJ...664..675E} to account for the effects
of survey geometry as well as redshift-space distortions and validate it on 160 LasDamas simulations. 
We demonstrate that reconstruction sharpens the BAO feature in the angle averaged
galaxy correlation function, reducing the nonlinear smoothing scale
$\Sigma_{\rm nl}$ from 8.1 Mpc/$h$ to 4.4 Mpc/$h$. Reconstruction also 
significantly reduces the effects of redshift-space distortions at the BAO scale, isotropizing 
the correlation function. This sharpened BAO feature yields an unbiased distance
estimate ($< 0.2\%$) and reduces the scatter from $3.3\%$ to $2.1\%$. We demonstrate the robustness of these
results to the various reconstruction parameters, including the smoothing scale, the galaxy bias and the 
linear growth rate. Applying this reconstruction algorithm to the Sloan Digital Sky Survey (SDSS) Luminous
Red Galaxy (LRG) Data Release 7 (DR7) sample improves the significance of the BAO feature 
in these data from $3.3\sigma$ for the
unreconstructed correlation function, to $4.2\sigma$ after reconstruction.
We estimate a relative distance scale $D_{V}/r_{s}$ to $z=0.35$ 
of $8.88~\pm~0.17$, where $r_{s}$ is the
sound horizon and $D_{V} \equiv (D_{A}^2 H^{-1} )^{1/3}$ is a combination of the angular diameter 
distance $D_A$ and Hubble parameter $H$. Assuming a sound horizon of 154.25 Mpc, this translates into 
a distance measurement $D_{V} (z=0.35) = 1.356~\pm~0.025\,{\rm Gpc}$. We find that reconstruction reduces
the distance
error in the DR7 sample from 3.5\% to 1.9\%, equivalent to a survey with three times the volume of SDSS. 
\end{abstract}

\section{Introduction}

The baryon acoustic oscillation method (hereafter BAO)
(see \citealt{2012arXiv1201.2434W} for a review) 
is a geometrical probe of the expansion 
rate of the Universe. Sound waves in the baryon-photon plasma are frozen as density fluctuations at
recombination, with a characteristic scale set by the 
sound horizon \citep{1966JETP...22..241S,1970Ap&SS...7....3S,1970ApJ...162..815P,1987MNRAS.226..655B,1984ApJ...285L..45B,
1996ApJ...471..542H,1997Natur.386...37H,1998ApJ...496..605E}. 
These sound waves manifest themselves as a peak
in the matter (and therefore galaxy) correlation function at the scale of the sound horizon 
($\sim\!150$ Mpc for current concordance cosmologies) or equivalently as a series of oscillations 
in the power spectrum. Since the sound horizon is precisely calibrated by cosmic microwave
background measurements, BAO measurements may be used as a standard ruler 
\citep{1997PhRvL..79.3806T,1998ApJ...495...29G,1998ApJ...504L..57E,1999MNRAS.304...75E}, 
mapping the angular
diameter distance (with the ruler aligned perpendicular to the line of sight) and the Hubble 
parameter (with the ruler parallel to the line of sight) as a function of redshift \citep{2003ApJ...594..665B,2003PhRvD..68f3004H,2003PhRvD..68h3504L,2003ApJ...598..720S}. 
There have
now been multiple detections of the BAO feature in galaxy surveys 
\citep{2005MNRAS.362..505C, 2005ApJ...633..560E, 2007MNRAS.374.1527B, 
2007MNRAS.378..852P, 2007MNRAS.381.1053P, 2009MNRAS.399..801G, 
2009MNRAS.399.1663G, 2010ApJ...710.1444K, 2010MNRAS.401.2148P, 
2010MNRAS.404...60R, 2011arXiv1102.2251C, 2011arXiv1108.2635B, 
2011MNRAS.415.2892B, 2011MNRAS.416.3017B, 2011MNRAS.416.3033S, 
2011MNRAS.tmp.1598B,2012arXiv1201.2137H,2012arXiv1201.2172S}
and current \citep{2009astro2010S.314S, 2010MNRAS.401.1429D, 2008ASPC..399..115H}
and planned \citep{2011arXiv1106.1706S, 2011arXiv1110.3193L}
BAO surveys are now a mainstay of experimental dark energy programs. 

A key feature of the BAO method is the fact that the sound horizon is much larger than the 
scales relevant for nonlinear evolution and galaxy formation. This scale separation protects the 
BAO feature from large corrections due to these effects and therefore from systematic errors. 
There is now a considerable literature quantitatively exploring these effects both using
perturbative techniques and simulations 
\citep{1999MNRAS.304..851M,2003ApJ...598..720S, 2006ApJ...651..619J, 2007APh....26..351H, 
2007MNRAS.375.1329G, 2007ApJ...664..660E, 2008MNRAS.383..755A, 
2008PhRvD..77b3533C, 2008PhRvD..77d3525S, 2008ApJ...686...13S, 
2009PhRvD..80f3508P, 2010ApJ...720.1650S, 2011ApJ...734...94M}
and consensus
that systematic effects that might bias distance estimates with the standard ruler are indeed small. 

The dominant effect of the nonlinear evolution of the density field is to smooth the BAO feature 
in the correlation function. This is equivalent to suppressing the oscillations in the power spectrum.
While this smoothing does not bias the distance measurements, it does reduce the constrast in the BAO feature 
and increases the distance errors. This smoothing is well understood
\citep{2007ApJ...664..660E, 2008PhRvD..77b3533C, 2009PhRvD..80f3508P, 2008PhRvD..77f3530M, 
2008PhRvD..78h3519M,2010ApJ...720.1650S}  
and is physically caused by large-scale ($\sim\!20 {\rm Mpc}/h$) flows.
This realization led \cite{2007ApJ...664..675E} to suggest that this smoothing of the BAO 
feature may be reversed, a process commonly referred to as ``reconstruction''.  They provide 
a simple prescription for this process that has been shown to sharpen the BAO feature
and improve distance constraints 
\citep{2009PhRvD..79f3523P, 2009PhRvD..80l3501N,2010ApJ...720.1650S, 2011ApJ...734...94M}. 
It is important to emphasize here that this is not a deconvolution but 
rather uses information beyond the two-point function that exists in the density field. 

Although reconstruction has been studied with simulations in the past, the present paper represents the 
first application to data. The Sloan Digital Sky Survey (SDSS) Luminous Red Galaxy (LRG) sample 
represents the current state of the art in low-redshift BAO measurements and is a natural sample to implement 
reconstruction on. 
This sample was analyzed by \citet{2010MNRAS.401.2148P} using the power spectrum and \citet{2010ApJ...710.1444K}
using the correlation function, who report a $\sim\!3.5\%$ distance measurement for the LRG sample alone, 
and a $\sim\!2.7\%$ measurement when combined with a lower redshift sample of galaxies from the SDSS. 
This work is the natural extension of these previous results.
In addition to improving the distance constraints from this sample, the performance 
of reconstruction on the SDSS has important implications for the expected performance of future
surveys, most of which assume some level of reconstruction.

This is the first in a sequence of three papers. This paper 
describes the details of the 
reconstruction algorithm used, tests it on simulated data, and then presents the results for the DR7 data. The second
paper in this series \citep[][hereafter Paper II]{Paper2} describes how we robustly fit the galaxy correlation 
function to obtain our distance measurements. The third paper \citep[][hereafter Paper III]{Paper3} 
presents the cosmological implications of these measurements. 

This paper is structured as follows: \S~\ref{sec:methods} introduces the basic principle behind reconstruction
and then presents a detailed description of the algorithm implemented in this work. We describe the data and 
simulations used in \S~\ref{sec:data}. We discuss the impact of reconstruction on the BAO feature and
the derived distances on simulated data in \S~\ref{sec:sims}; we then apply it to the data in \S~\ref{sec:dr7}.
We summarize our conclusions in \S~\ref{sec:discuss}.

\section{Methods}
\label{sec:methods}

We describe the various algorithms used in the reconstruction of the BAO feature below. We start by describing the
physical basis for reconstruction and outline the reconstruction algorithm. We then describe our approach to 
dealing with survey boundaries, as well as our procedure for estimating a distance scale from a correlation 
function.

\subsection{Understanding Reconstruction}

\begin{figure*}
  \includegraphics[width=3in]{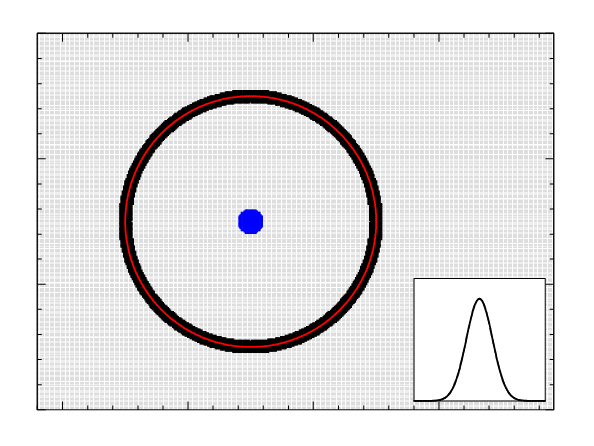}
  \includegraphics[width=3in]{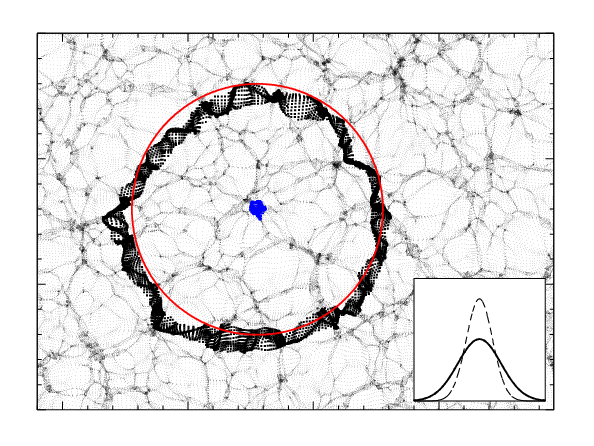}
  \includegraphics[width=3in]{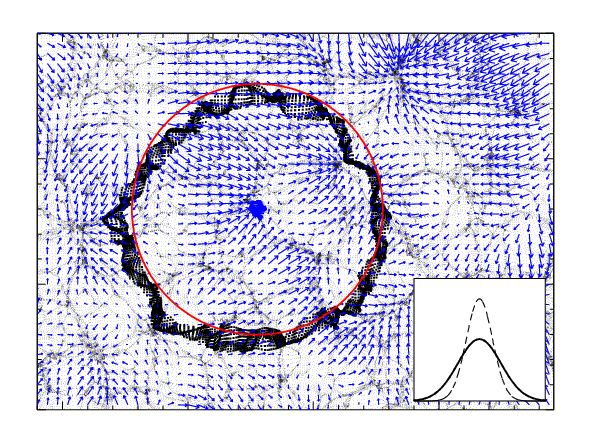}
  \includegraphics[width=3in]{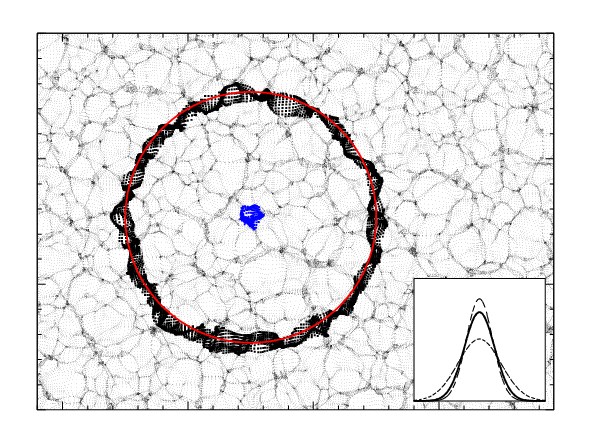}
  \caption{A pictoral explanation of how density-field reconstruction can improve
  the acoustic scale measurement.
  In each panel, we show a thin slice of a simulated cosmological density field.
  ({\it top left}) In the early universe, the initial densities are
  very smooth.  We mark the acoustic feature with a ring of 150 Mpc
  radius from the central points.  A Gaussian with the same rms width
  as the radial distribution of the black points from the centroid of
  the blue points is shown in the inset.
  ({\it top right}) We evolve the particles to the present day, here by
  the Zel'dovich approximation \citep{1970A&A.....5...84Z}.  The red circle shows
  the initial radius of the ring, centered on the current centroid of 
  the blue points.  The large-scale velocity field has caused the black
  points to spread out; this causes the acoustic feature to be broader.
  The inset shows the current rms radius of the black points relative
  to the centroid of the blue points (solid line)
  compared to the initial rms (dashed line).
  ({\it bottom left}) As before, but overplotted with the Lagrangian
  displacement field, smoothed by a 10$h^{-1}$ Mpc Gaussian filter.
  The concept of reconstruction is to estimate this displacement field
  from the final density field and then move the particles back to
  their initial positions.
  ({\it bottom right}) We displace the present-day position of the
  particles by the opposite of the displacement field in the previous
  panel.  Because of the smoothing of the displacement field, the
  result is not uniform.  However, the acoustic ring has been moved
  substantially closer to the red circle.  The inset shows that the
  new rms radius of the black points (solid), compared to the initial
  width (long-dashed) and the uncorrected present-day width (short-dashed).  The narrower peak
  will make it easier to measure the acoustic scale.  Note that the 
  algorithm applied to the data is more complex than was just described, but
  this figure illustrates the basic opportunity of reconstruction.}
  \label{fig:recon_movie}
\end{figure*}

Figure~\ref{fig:recon_movie} highlights the key aspects of reconstruction. As was first emphasized by
\citet{2007ApJ...664..660E}, the erasure of the BAO feature can be physically traced to the pairwise
relative velocities of particles separated by $\sim\!100$ Mpc$/h$. 
Figure~\ref{fig:recon_movie} highlights this, showing a slice through an N-body simulation, with tracer 
particles representing the BAO feature highlighted. The 
dominant smoothing of the BAO feature is due to the coherent flows that form the large-scale structure, 
not the random motions of particles within gravitational structures. This is the key 
insight behind reconstruction - the same galaxy surveys used to detect the BAO feature also map the
cosmic structure responsible for its erasure. One can therefore use these same surveys to infer the large
scale flow field and partially undo the smoothing of the BAO feature. 

Figure~\ref{fig:recon_movie} also makes it clear that reconstruction is actually working at the
level of the density field,  using information beyond what exists simply in the 
two-point statistics of the field. The gains of reconstruction therefore {\it cannot} be achieved by simply 
forward modeling the correlation function into the nonlinear regime. While one might attempt to recover
this information by considering higher order statistics of the density field, this is an 
awkward encoding of the information required.

The steps outlined in the figures are a simplification of the reconstruction algorithm we use; we
describe the detailed algorithm next.

\subsection{A Reconstruction Algorithm}

We implement an extended version of the reconstruction algorithm of \cite{2007ApJ...664..675E}. The 
theoretical underpinnings of this algorithm have been described in \cite{2009PhRvD..79f3523P} and 
\cite{2009PhRvD..80l3501N},
and the algorithm has been validated against several different suites of simulations 
\citep{2010ApJ...720.1650S, 2011ApJ...734...94M}.
We outline the steps of this algorithm below and discuss details specific to our implementation 
in subsequent subsections.

\begin{enumerate}
  \item Estimate the unreconstructed power spectrum $P(k)$ or correlation function $\xi(r)$. 
  \item Estimate the galaxy bias $b$ and the linear growth rate, 
  $f \equiv d \ln D/d \ln a \sim\! \Omega_{M}^{0.55}$
  \citep{1992ARA&A..30..499C, 2005PhRvD..72d3529L}, where $D(a)$ is the linear growth function as a function 
    of scale factor $a$ and $\Omega_{M}$ is the matter density relative to the critical density.
  \item Embed the survey into a larger volume, chosen such that the boundaries of this larger
  volume are sufficiently separated from the survey. 
  \item Gaussian smooth the density field.
  \item Generate a constrained Gaussian realization that matches the observed density 
  and interpolates over masked and unobserved regions (\S\ref{sec:hoffrib}).
  \item Estimate the displacement field $\mathbf{\Psi}$ within the Zel'dovich approximation (\S\ref{sec:pot}).
  \item Shift the galaxies by $-\mathbf{\Psi}$. 
    Since linear redshift-space distortions arise from the same velocity field, we  
    shift the galaxies by an additional $-f(\mathbf{\Psi} \cdot \hat{\mathbf{s}}) \hat{\mathbf{s}}$
    (where $\hat{\mathbf{s}}$ is the radial direction). In the limit of linear theory (i.e. large scales), 
    this term exactly removes redshift-space distortions \citep{1987MNRAS.227....1K,1998ASSL..231..185H,2004PhRvD..70h3007S}. 
    Denote these points by $D$.
  \item Construct a sample of points randomly distributed according to the angular and radial selection
  function and shift them by $-\mathbf{\Psi}$. Note that we do not correct these for redshift-space distortions. 
  Denote these points by $S$. 
  \item The reconstructed correlation function $\xi$ is then given by the Landy-Szalay \citep{1993ApJ...412...64L}
  estimator, 
  \begin{equation}
    \xi = \frac{DD - 2DS + SS}{RR}
  \end{equation}
  where $DD$ etc are the number of pairs at a given separation between various sets of points. The 
  random points $R$ are  distributed randomly according to the angular and radial selection
  functions; these are assumed to be different from those to generate $S$. We weight the points by 
  an approximate minimum variance weight \citep{1994ApJ...426...23F}, 
  \begin{equation}
    w_{i} = \frac{1}{1 + \bar{n}(z_{i}) P(k_0)} 
  \end{equation}
  where $\bar{n}$ is the redshift distribution at the galaxy redshift $z_{i}$ and $P(k_0) = 40000 ({\rm Mpc}/h)^3$
  is the power spectrum approximately at the BAO scale for SDSS LRGs.
\end{enumerate}

\subsection{Generating Constrained Realizations}
\label{sec:hoffrib}

Since the gravitational potential (and therefore, the displacement) depends non-locally on the matter overdensity,
it is sensitive to regions of space either masked out by the survey or not surveyed at all.
To handle this complication, we embed the survey into a larger region where the boundaries
of this larger region are sufficiently far from the true survey (in what follows below, we pad by 
200 Mpc/$h$). Since the density 
field on large scales can be approximated by a Gaussian density field, the problem of ``filling'' in 
the missing regions is then equivalent to the problem of generating constrained realizations of 
a Gaussian density field (\citealt{1991ApJ...380L...5H}, see also \citealt{1995ApJ...449..446Z}, who demonstrate that the previous
algorithm is equivalent to a Weiner filtering of the density field). Strictly speaking, one should 
marginalize over the ensemble of such realizations. However, as we demonstrate below, our results are 
insensitive to the details of this implementation. Therefore, for simplicity, we consider only a single 
realization.

We start by organizing the observed and constrained realization density fields into an $N_{\rm obs}$ element vector 
$\bd$ and an $N_{\rm embed}$ element vector $\tilde{\bd}$, related by 
the trivial projection ${\bm P}$ on to the observed points, $\bd = {\bm P}\tilde{\bd}$. 
The Hoffman-Ribak algorithm is then 
\begin{equation}
  \tilde{\bd} = \tilde{\bd}_{U} + \tilde{\bm C} {\bm C}^{-1} \left(\bd - {\bm P}\tilde{\bd}_{U} \right)
\end{equation}
where $\tilde{\bd}_{U}$ is an unconstrained Gaussian realization, with an assumed power spectrum $P(k)$. 
The covariance matrices ${\bm C}$ and $\tilde{\bm C}$ are defined by the correlation function between
pairs of pixels,
\begin{equation}
  C_{ij} \equiv \langle \delta({\bm r}_{i}) \delta({\bm r}_{j}) \rangle \,\,, 
\end{equation}
where the correlation function is just the Fourier transform of the power spectrum. Note that 
${\bm C}$ is an $N_{\rm obs} \times N_{\rm obs}$ matrix and operates only on the observed pixels, while 
$\tilde{\bm C}$ has dimension $N_{\rm embed} \times N_{\rm obs}$ and relates the observations and
constrained realization.

While this algorithm is straightforward in principle, a number of comments are in order regarding 
its implementation. The first concerns our choice of power spectrum for generating the constrained 
realization and covariance matrices. {\it Unlike} the algebraicly similar problem  of power spectrum estimation, we do have 
a measurement of the power spectrum, which we use. A related issue is that in redshift space, the
prior correlation function is not isotropic and translation-invariant. We ignore this subtlety for simplicity and 
use an isotropic correlation function. We demonstrate later that, for the SDSS DR7 geometry, the reconstructed 
correlation function is robust to these choices. Finally, we note that our prior power spectrum 
also includes a white noise component of amplitude $\bar{n}^{-1}$.

The second issue is computational. The dimensions of $\bd$ and $\tilde{\bd}$ are $N\sim\! {\cal O}(10^{9})$, making
direct manipulation impossible. To proceed, we use the fact that multiplication by ${\bm C}$ is equivalent to 
a convolution by the correlation function \citep{2003NewA....8..581P}. Since we have ignored the angular 
dependence of the power spectrum, we can implement this in ${\cal O}(N \log N)$ time using FFTs. The matrix inverse
operations are implemented using a preconditioned conjugate gradient algorithm \citep{1992nrca.book.....P}  with 
the preconditioner being the convolution by the Fourier transform of $1/P(k)$. For an unmasked survey, 
this preconditioner is the exact inverse. The above allows us to generate constrained realizations in 
${\cal O}(10)$ iterations. 

\subsection{Solving for the Displacement}
\label{sec:pot}

To linear order, the displacement ${\bm \Psi}$ can be related to the density in redshift space by \citep{1994ApJ...421L...1N}
\begin{equation}
  \nabla \cdot {\bm F} = 
    \nabla \cdot {\bm \Psi} + f \nabla \cdot (\Psi_{s} \hat{\bm s}) = -\frac{\delta_{\rm gal}}{b}\,,  
\end{equation}
where $\Psi_{s} \equiv \mathbf{\Psi} \cdot \hat{s}$ is the displacement in the redshift direction and 
$\delta_{\rm gal}$ is the galaxy overdensity. 
Assuming the $\mathbf{\Psi}$ is irrotational, we write $\mathbf{\Psi} = \nabla \phi$ and solve for 
the scalar valued $\phi$. The resulting equation resembles Poisson's equation with an additional term 
for redshift-space distortions. However, the redshift-space term breaks the translational invariance 
of the problem and prevents us from solving this with FFTs\footnote{Note that in the plane-parallel
approximation, translational invariance is restored.}. We solve this equation by converting all
the derivatives to their finite difference counterparts and solve the resulting linear equation. We
implement this using the parallel GMRES algorithm in the 
PETSC \citep{petsc-web-page, petsc-user-ref, petsc-efficient} toolkit.
Having computed $\phi$, we obtain the displacement field by finite differences.
An advantage of this formulation is that the algorithm is easily extended to non-Cartesian 
coordinate systems, although we do not use this feature in this work.

\subsection{Fitting the Acoustic Feature}

We briefly describe our fitting procedure below; a detailed description and tests of this procedure in 
in Paper II.
We define a fiducial fitting model
\begin{equation}
\xi^{\rm fit}(r) = B^2\xi_m(\alpha r)+A(r)
\label{eqn:fform}
\end{equation}
where
\begin{equation} 
\xi_m(r) = \int \frac{k^2dk}{2\pi^2}P_m(k)j_0(kr) e^{-k^2a^2},
\end{equation}
and
\begin{equation}
A(r) = \frac{a_1}{r^2} + \frac{a_2}{r} + a_3.
\label{eqn:aform}
\end{equation}
Our template is determined by interpolating between the linear theory power spectrum 
and one with the BAO feature erased \citep{2007ApJ...664..660E} 
\begin{equation}
P_m(k) = [P_{\rm lin}(k)-P_{\rm smooth}(k)] e^{-k^2\Sigma_{\rm nl}^2/2}+P_{\rm smooth}(k).
\label{eqn:template}
\end{equation}
For convenience, we choose to normalize this template to the observed correlation
function at $r=50$ Mpc/$h$; this ensures $B^2 \sim\! 1$. 
The $\Sigma_{\rm nl}$
parameter smooths the BAO feature, 
modeling the degradation due to non-linear structure growth. The
$e^{-k^2a^2}$ term is used to damp the oscillatory transform
kernel $j_0(kr)$ at high-$k$ to induce better numerical convergence in
the integration. The $A(r)$ term, with the associated $a_{1,2,3}$ nuisance parameters, 
is used to help marginalize out the unmodeled 
broadband signal in the correlation function. This broadband signal 
includes redshift-space distortions,
scale-dependent bias and any errors we make in our assumption of the model
cosmology which might bias 
the acoustic peak. 

Our distance constraints are captured by 
the scale dilation parameter $\alpha$ which represents the
shift in the acoustic peak. An $\alpha>1$
indicates a shift towards smaller scales and an $\alpha<1$ indicates a
shift towards larger scales.

We obtain the best-fit value of $\alpha$ by minimizing the $\chi^2$
goodness-of-fit indicator 

In order to compute the likelihood function $p(\alpha)$ given the measured 
correlation function, we start with 
\begin{equation}
\chi^2(\alpha) = [\vec{d}-\vec{m}(\alpha)]^TC^{-1}[\vec{d}-\vec{m}(\alpha)]
\end{equation}
where $\vec{d}$ is the measured correlation function
and $\vec{m}(\alpha)$ is the best-fit model at each $\alpha$. We use
a fiducial fitting range of 30-200$h^{-1}\rm{Mpc}$. $C$ is a modified
Gaussian covariance matrix described in detail in Paper II. We minimize this
function for a grid of fixed values of $\alpha$; this step is exactly equivalent
to marginalizing over the linear parameters $a_{i}$, and is a good approximation 
for the $B^2$ term as well. The likelihood distribution of $\alpha$ is then simply
\begin{equation}
  p(\alpha_{i})  \propto e^{-\chi_{i}^2/2} 
\end{equation}
where $\chi_{i}$ is the $\chi^2$ minimum at $\alpha_i$ and the proportionality
constant is determined by ensuring that the probability integrates to one. In addition, 
we impose a 15\% Gaussian prior on $\log(\alpha)$. While this prior does not affect
the core of the probability distribution, it does suppress values of $\alpha \ll 1$ that correspond
to the BAO scale being shifted to very large scales, reflecting our current state of knowledge
about the background cosmology. 

This probability distribution captures all the distance information of this data and is what 
we use when estimating cosmological parameters in Paper III. It is however convenient to 
summarize this information. We do so throughout this paper by reporting the mean $\langle \alpha \rangle$
and standard deviation $\sigma_{\alpha}$ for the data and each individual mock catalog, where
$\sigma_{\alpha}^2 = \langle \alpha^2 \rangle - \langle \alpha \rangle^2$ and 
\begin{equation}
  \langle \alpha^{n} \rangle = \int d\alpha\,p(\alpha) \alpha^n \,\,.
\end{equation}
As we see below for the data and is discussed in greater detail in Paper II, $p(\alpha)$ is well approximated
by a Gaussian and therefore completely characterized by the mean and standard deviation. 
It is also worth emphasizing that this procedure yields an independent measurement of the distance error
for every mock, capturing the effects of sample fluctuations. Our analyses must therefore not just 
explore the distribution of $\alpha$ but also of $\sigma_{\alpha}$ as well (see below).

\section{Data and Simulations}
\label{sec:data}

\subsection{The LRG sample}

The Sloan Digital Sky Survey \citep{2000AJ....120.1579Y}  has imaged $\sim\!$ 10,000 deg$^2$ of the sky, 
and obtained spectra of nearly a million of the detected objects. The imaging was carried out
by drift scanning the sky in photometric conditions \citep{2001AJ....122.2129H} in the $ugriz$ bands \citep{1996AJ....111.1748F,2002AJ....123.2121S}
with the Apache Point 2.5m telescope \citep{2006AJ....131.2332G} using a specially designed wide field camera 
\citep{1998AJ....116.3040G}. These data were processed by completely automated pipelines that detect and 
measure the photometric properties of the objects and astrometrically and photometrically calibrate
these observations 
\citep{2003AJ....125.1559P,2004AN....325..583I,2006AN....327..821T,2008ApJ...674.1217P}. 
Subsamples from the resulting photometric samples were 
selected \citep{2002AJ....124.1810S,2001AJ....122.2267E} for spectroscopy using a 640 fiber spectrograph. The SDSS has had three phases:
SDSS-I and SDSS-II completed the observations described above in 2009 and the data were released 
in a series of seven data releases, with the SDSS Data Release 7 \citep[DR7]{2009ApJS..182..543A} being the final data release. 
The third phase of SDSS \citep[SDSS-III]{2011AJ....142...72E} started taking data in 2009 and includes the Baryon
Oscillation Spectroscopic Survey \citep[BOSS]{2009astro2010S.314S} as part of its science goals.

\begin{figure}
  \includegraphics[width=3in]{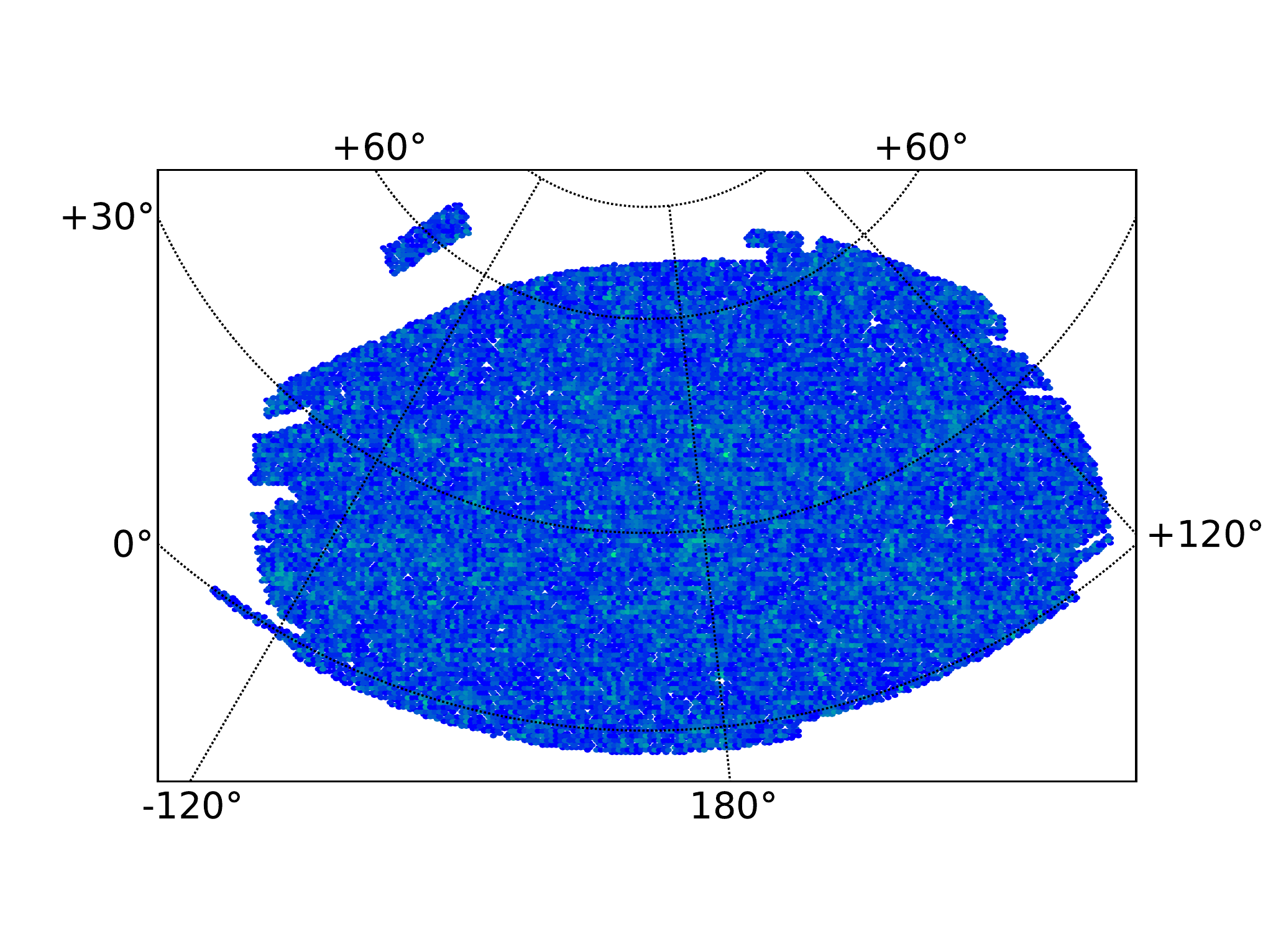}
  \caption{The footprint of the DR7 LRG sample used in this paper, plotted in equatorial coordinates and 
  an Albers equal area projection. The area covered is 7189 deg$^2$.}
  \label{fig:dr7_radec}
\end{figure}

The sample that we consider for this analysis is the LRG sample. The 
motivation and selection of this sample is described in detail in \citet{2001AJ....122.2267E} and we
refer the interested reader to the description there. These galaxies are very luminous
and therefore can probe cosmologically interesting volumes. Furthermore, they are generally
old stellar systems with very uniform spectral energy distributions, characterized by a strong break
at 4000\AA. This gives these galaxies a distinct color-flux-redshift relation, which allows 
them to be uniformly selected over a broad redshift range. The LRG samples have been used for a 
number of cosmological analyses in the SDSS, including the BAO detection in the SDSS
\citep{
2005ApJ...633..560E, 2007MNRAS.374.1527B, 
2007MNRAS.378..852P, 2007MNRAS.381.1053P, 2009MNRAS.399..801G, 
2009MNRAS.399.1663G, 2010ApJ...710.1444K, 2010MNRAS.401.2148P, 
2010MNRAS.404...60R, 2011arXiv1102.2251C, 2011MNRAS.416.3033S} 
We use a sample identical to that used in \citet{2010ApJ...710.1444K} and refer the reader there for 
a detailed description of its construction.

\begin{figure}
  \includegraphics[width=3in]{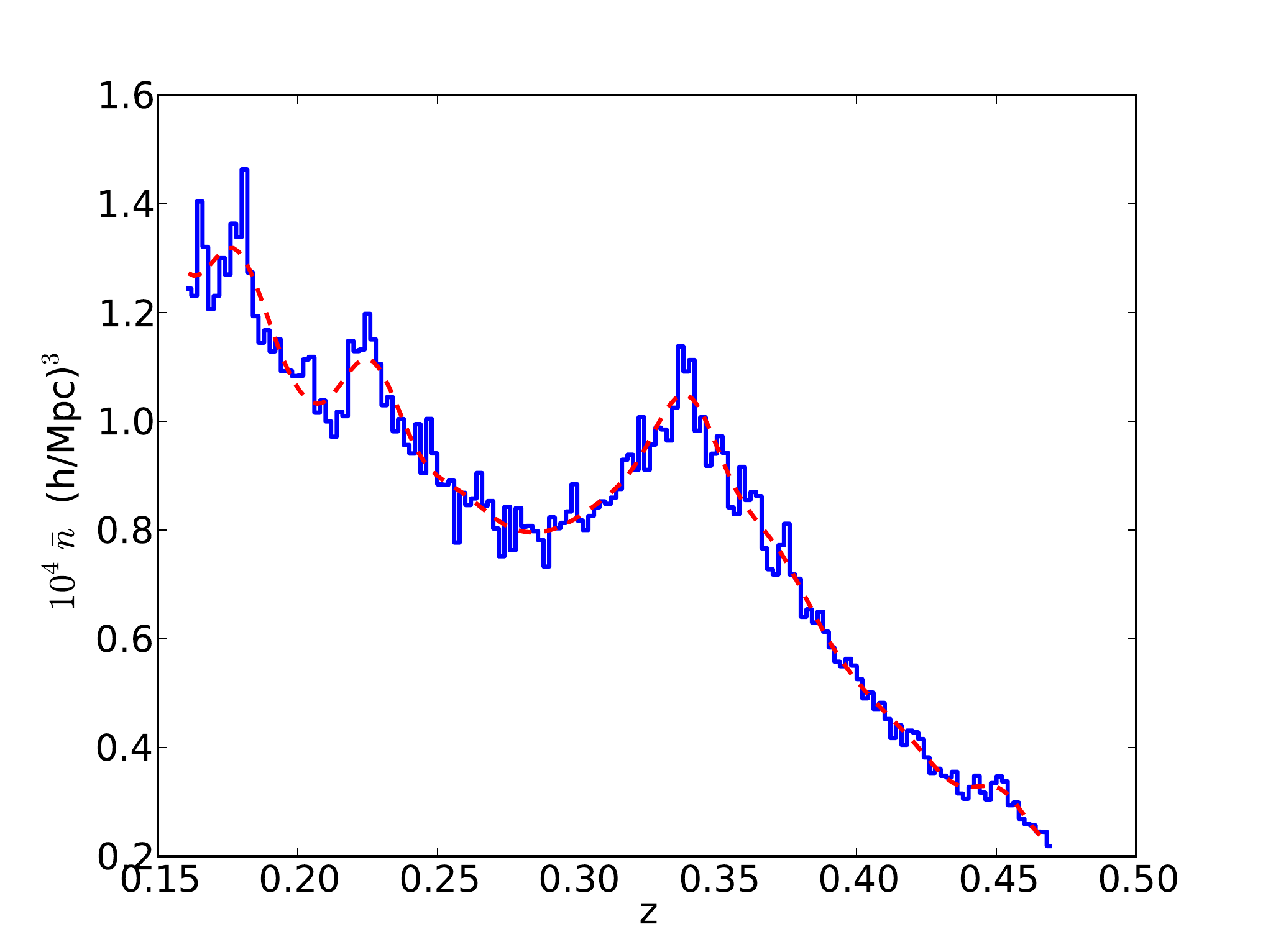}
  \caption{The redshift distribution of the DR7 LRG sample used in this paper. The dashed [red] line is a smooth
  fit to the redshift distribution used in the determination of the weights used in the correlation function.}
  \label{fig:dr7_nbar}
\end{figure}

Figure~\ref{fig:dr7_radec} shows the sky coverage of the sample that we consider. While the SDSS-II survey has 
data in both the Northern and Southern Galactic Caps, we focus on the contiguous footprint in the North, with
a total area of 7189 deg$^2$. Figure~\ref{fig:dr7_nbar} plots the number density of the LRG sample as function 
of redshift. We truncate the redshift distribution at $z=0.16$, since the color selection of the LRG sample
breaks down there. The redshift distribution is approximately constant out to a redshift of $z\sim\! 0.35$, 
falling off due to the magnitude limit of the sample beyond.

\subsection{LasDamas Mock Galaxy Catalogs}

We test our implementation of the reconstruction algorithm on mock galaxy 
catalogs created from the 
Large Suite of Dark Matter Simulations (LasDamas; McBride et al 2012, in prep).
Our goals are 
(a) to demonstrate that reconstruction yields an unbiased and improved distance scale measurement, 
(b) to test the robustness of the reconstruction algorithm to its input parameters, and 
(c) to tune these parameters in a ``blind'' manner.

We use the publicly available mock galaxy catalogues constructed by the 
LasDamas collaboration \footnote{http://lss.phy.vanderbilt.edu/lasdamas}, 
which are designed to simulate the SDSS LRG samples.  We chose the gamma release of 
\texttt{lrgFull} (in the LasDamas nomenclature) which best corresponds to our data sample.  
LasDamas assumes a flat $\Lambda$CDM cosmology, roughly consistent with WMAP5, with 
$\Omega_{\rm baryon}=0.04$, 
$\Omega_{\rm matter} = 0.25$, 
$\Omega_{\Lambda} = 0.75$, 
$h=0.7$, 
$n_{s}=1.0$, 
and $\sigma_{8} = 0.8$.
The LRG mocks were constructed from 40 ``Oriana'' N-body realizations, which were 
each run in a large cubical volume ($L = 2.4 \; h^{-1}$ Gpc) with $1280^3$ 
particles and initialized using second-order Lagrangian perturbation theory 
at $z=49$.  The mock galaxy catalogs were constructed by populating dark 
matter halos catalogs, where the halo occupation parameters were varied to 
match observed galaxy clustering measurements.  The mock galaxy catalogs 
modeled the realism of observed data by modeling redshift distortions, matching 
the angular selection function of the SDSS DR7 LRG sample and spanning the 
redshift range of $z=0.16-0.44$.  We make one modification to the public 
catalogs to better match our analysis: we downsample the radial selection 
function to match our specific LRG data.  This is necessary to properly model 
the galaxy numbers in the flux-limited region of the LRG selection at redshifts 
greater than $z=0.36$.  We make use of the catalogs covering only the 
northern galactic cap of the SDSS footprint ($7214.34$ sq.deg), which yield 
four mocks from each simulation for a total of 160 galaxy catalogs.  

\subsection{Fiducial Cosmologies}

We conclude this section by discussing two technical details --- our adopted definition of the sound horizon and 
the fiducial cosmology assumed in our analyses. We follow \cite{1998ApJ...496..605E} 
and assume the sound horizon specified
by Eq.6 of that paper and defer a comparison of the different choices to Paper III. 
We use two fiducial cosmologies in our analyses below; while both of these are flat $\Lambda$CDM cosmologies, 
they differ in their choices of parameters. For the LasDamas simulations, we use the cosmology assumed 
for the simulations - a baryon density of $\Omega_{\rm b} =  0.04$, a matter density of $\Omega_{\rm m} = 0.25$, 
and a Hubble constant of 70 km/s/Mpc ($h=0.7$). However, these choices differ significantly from the current best fit
to the cosmic microwave background data from the WMAP satellite 
\citep[][hereafter WMAP7]{2011ApJS..192...18K}, with
$\Omega_{\rm b} = 0.0457$, $\Omega_{\rm m} = 0.274$ and $h=0.702$.
In particular, the sound horizon assuming the WMAP7 cosmology is 152.76 Mpc, compared to 159.68 Mpc for the
LasDamas cosmology, a difference larger than our claimed statistical accuracy. While, as we discuss below, 
an incorrect cosmology does not bias our distance scale, it does change the errors on the distance scale. This 
makes it important to iterate and choose a cosmology close to the best fit in the analysis. As we
demonstrate below, the data are well fit by the WMAP7 cosmology and we use it as our fiducial model 
when analyzing the SDSS DR7 data.

\section{Reconstructing Simulations}
\label{sec:sims}

We start by discussing the impact of reconstruction on the LasDamas simulations.
The top left  panel is the unreconstructed real-space 
correlation function, with the BAO ring clearly visible and the correlation function independent of angle. 
Turning on redshift-space distortions destroys the isotropy of the correlation function, with
the maximal distortion, as expected, parallel to the line of sight. The distortions at small $r_\perp$ and 
$r_{||} < 20 {\rm Mpc}/h$ are 
due to virial motions inside motions inside halos, so called ``Fingers of God'' (FoG). 
The bottom panels show the 
correlation functions after reconstruction, assuming a 15 Mpc/$h$ smoothing length to estimate the displacement 
field and the true values of the logarithmic growth rate $f$ and the galaxy bias $b$. 
The most relevant feature for this paper is the enhanced BAO signal, apparent from the increased contrast
of the BAO ring.
Equally striking
is the restored isotropy of the redshift-space correlation function, demonstrating that reconstruction is correcting 
for the large-scale redshift-space distortions. The breakdown on small scales is due to a combination of the fact that 
the model for the displacement field is based on linear theory and that reconstruction is imperfect on these scales. 
We also note that the fingers of God become more prominent, highlighting the tendency of reconstruction to 
blow up collapsed objects. It is important to emphasize that both these effects are restricted to small scales
and have no effect of the acoustic scale.

\begin{figure*}
  \includegraphics[width=3in]{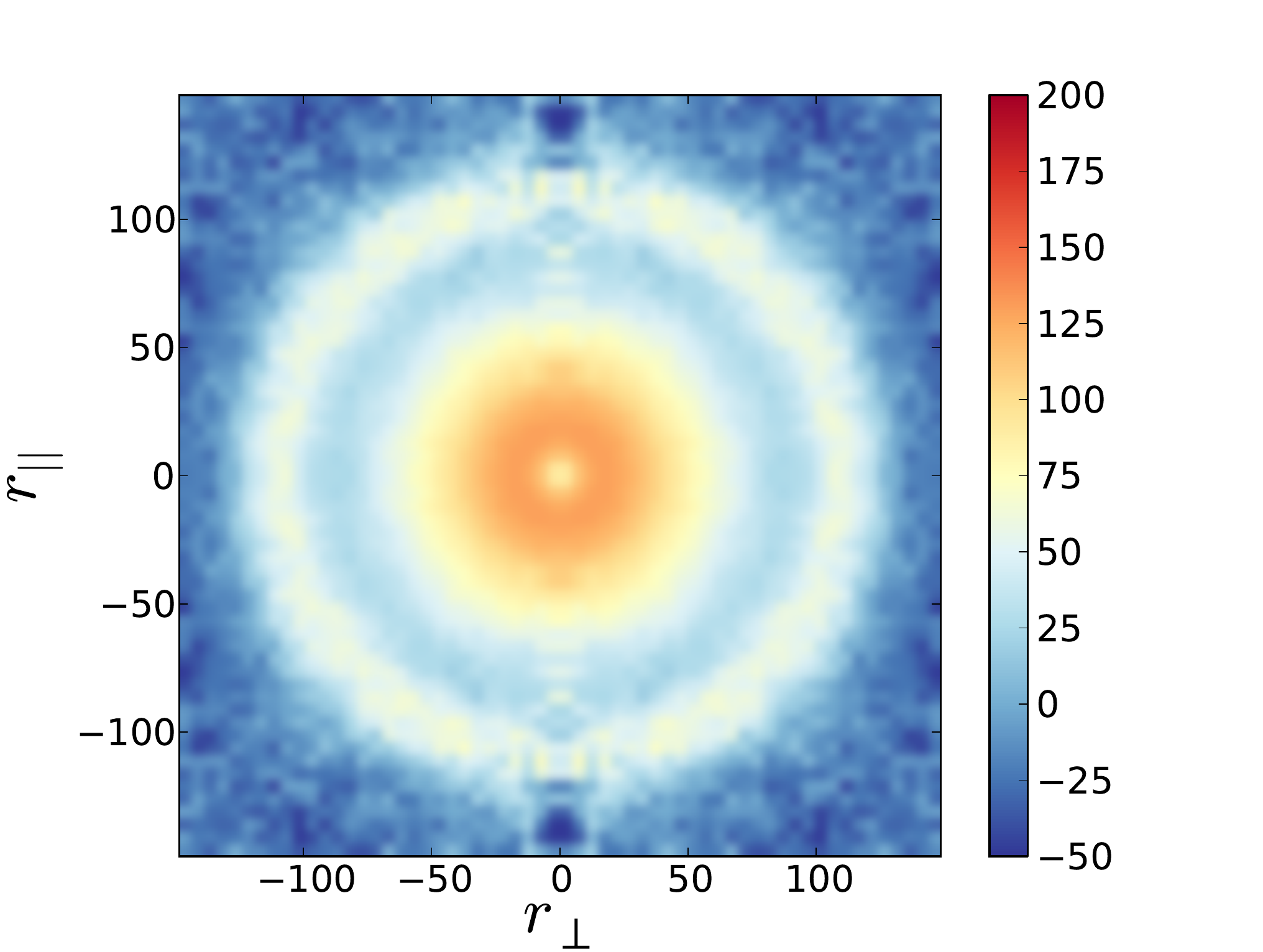}
  \includegraphics[width=3in]{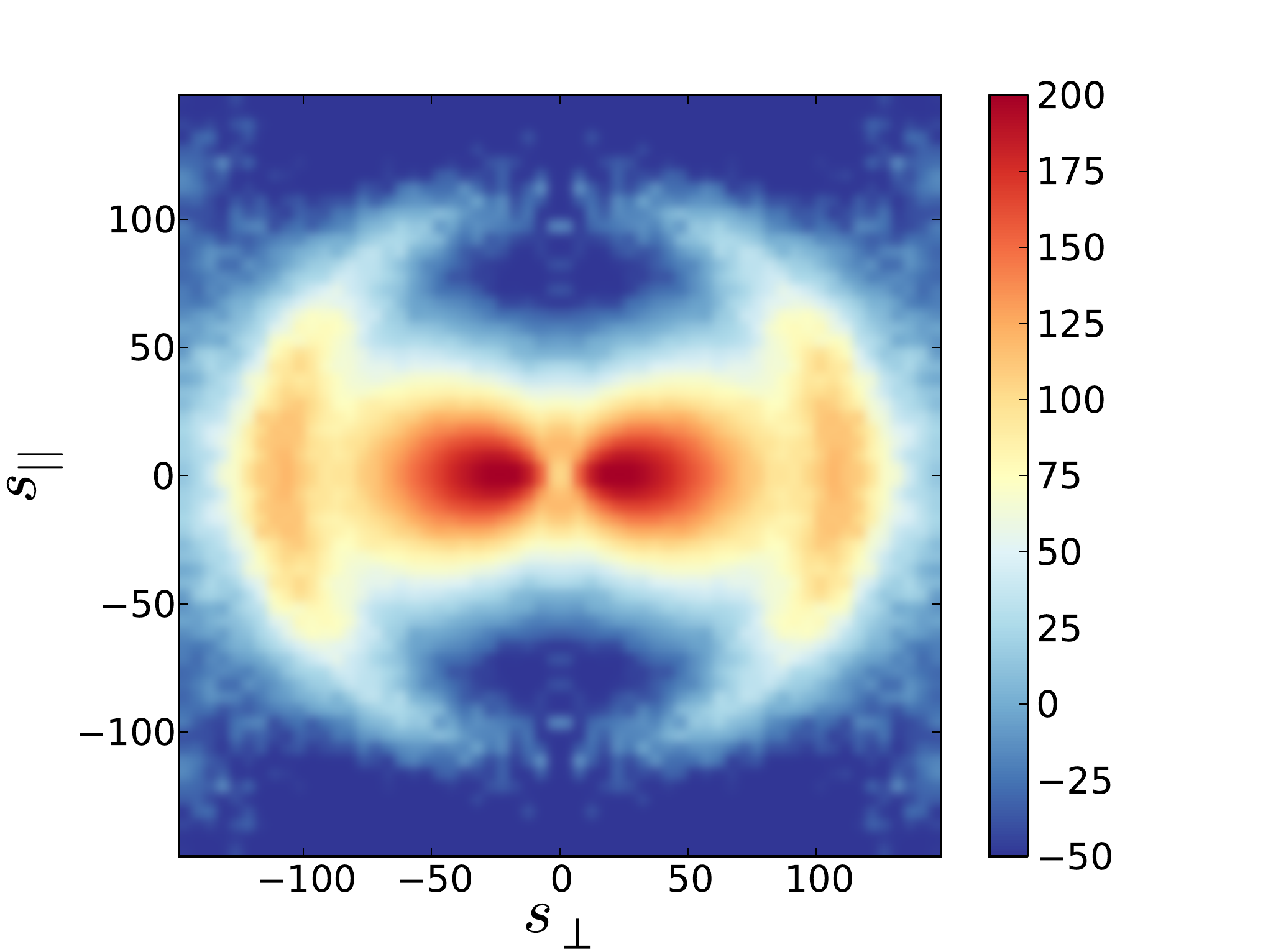}
  \includegraphics[width=3in]{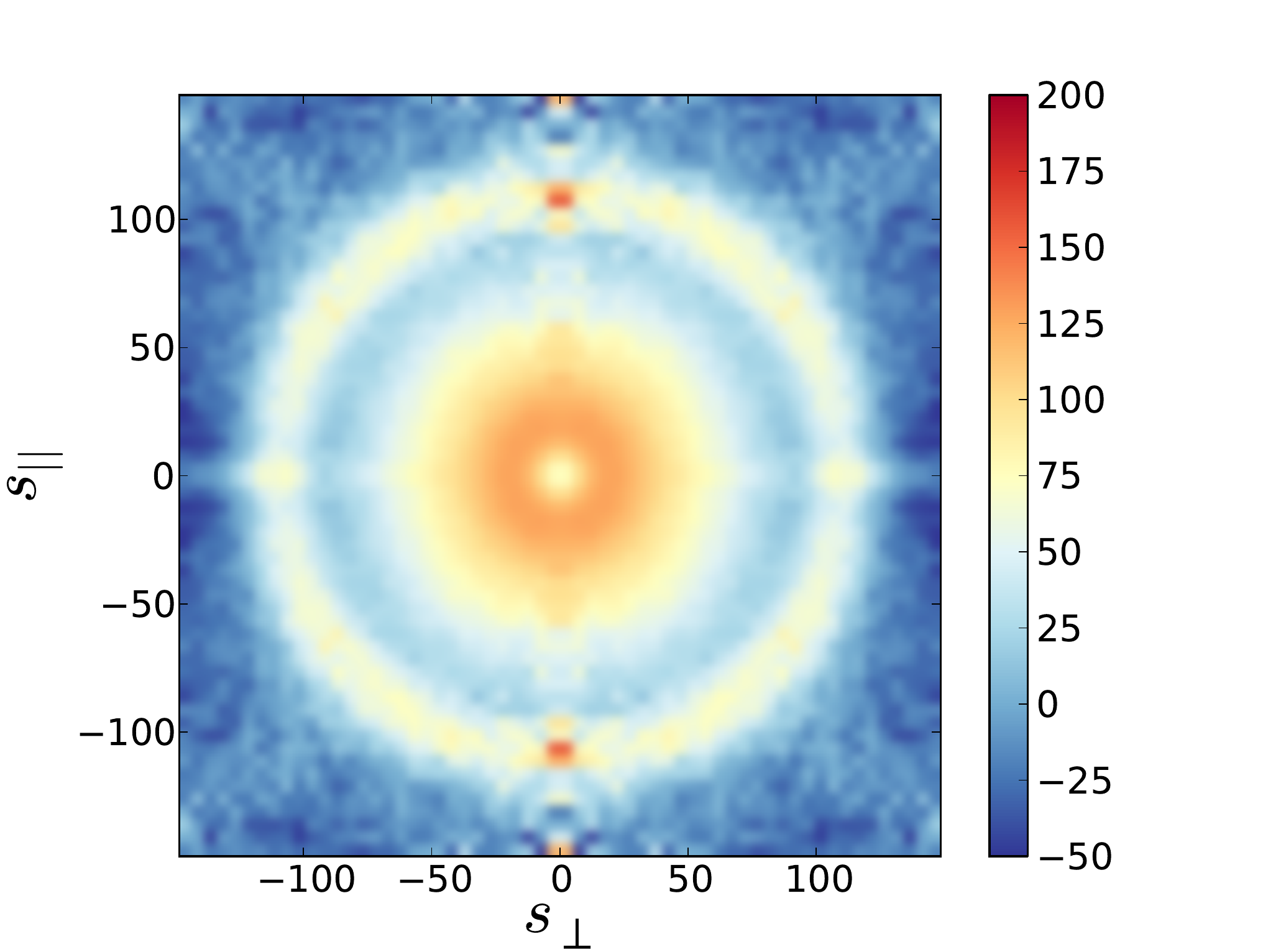}
  \includegraphics[width=3in]{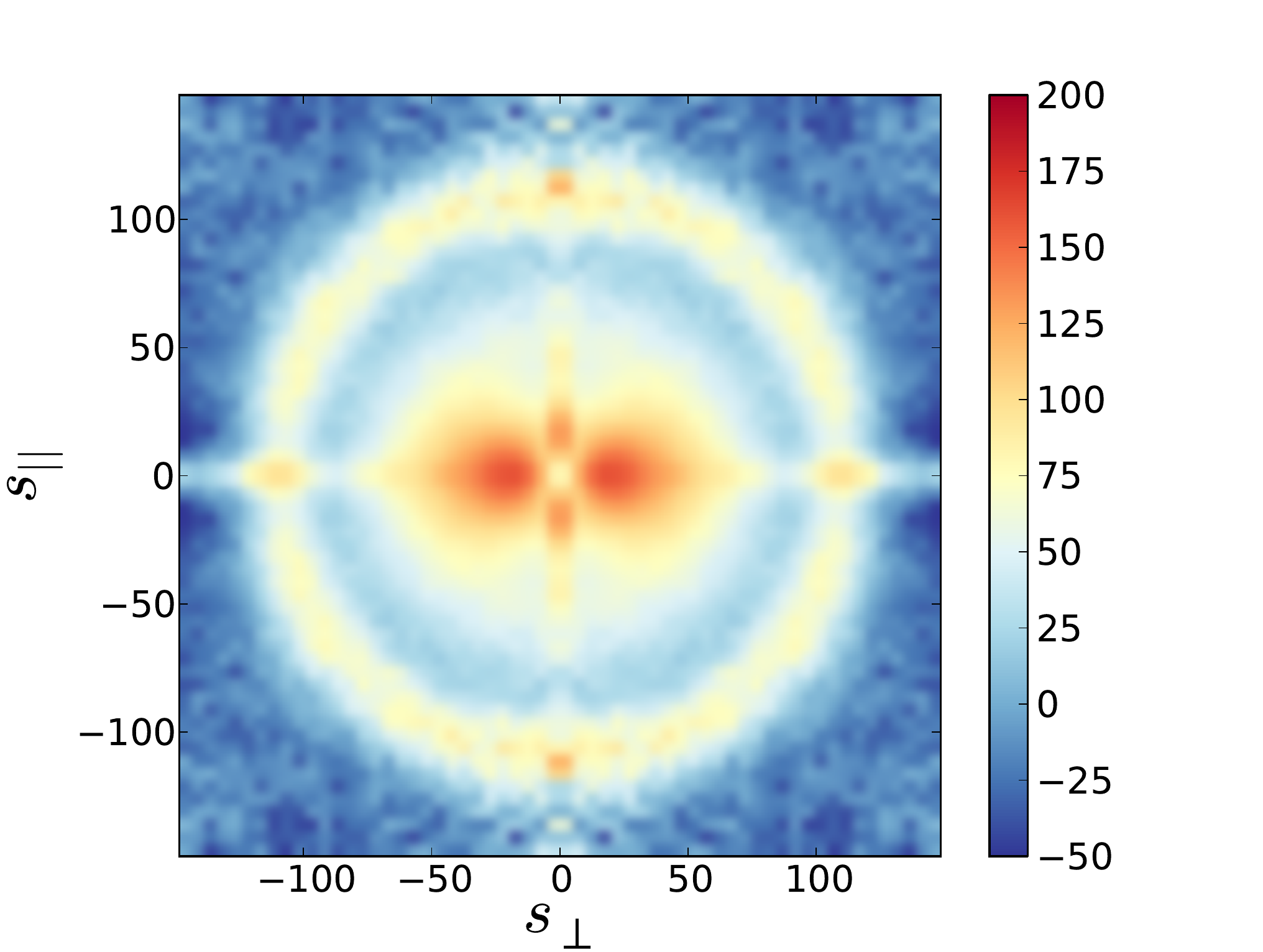}
  \caption{The LasDamas galaxy correlation function, averaged over the 160 simulations, as a function of the
  separation perpendicular ($\perp$) and parallel ($||$) to the line of sight. The correlation functions
  have been scaled by $r^2$ to highlight the BAO feature. 
  The top panels show the unreconstructed correlation functions, while the bottom panels show the reconstructed
  correlation functions; the left and right panels are real and redshift space respectively. The BAO feature is 
  visible as a ring at $\sim\! 110 {\rm Mpc}/h$ in the top left panel. Redshift space distortions 
  destroy the isotropy of the correlation function (top right). Reconstruction both sharpens the BAO feature (highlighted
  in the bottom left panel)
  and restores the isotropy (bottom right)  of the correlation function on the BAO scale.}
  \label{fig:wpav}
\end{figure*}

\begin{figure*}
  \includegraphics[width=3in]{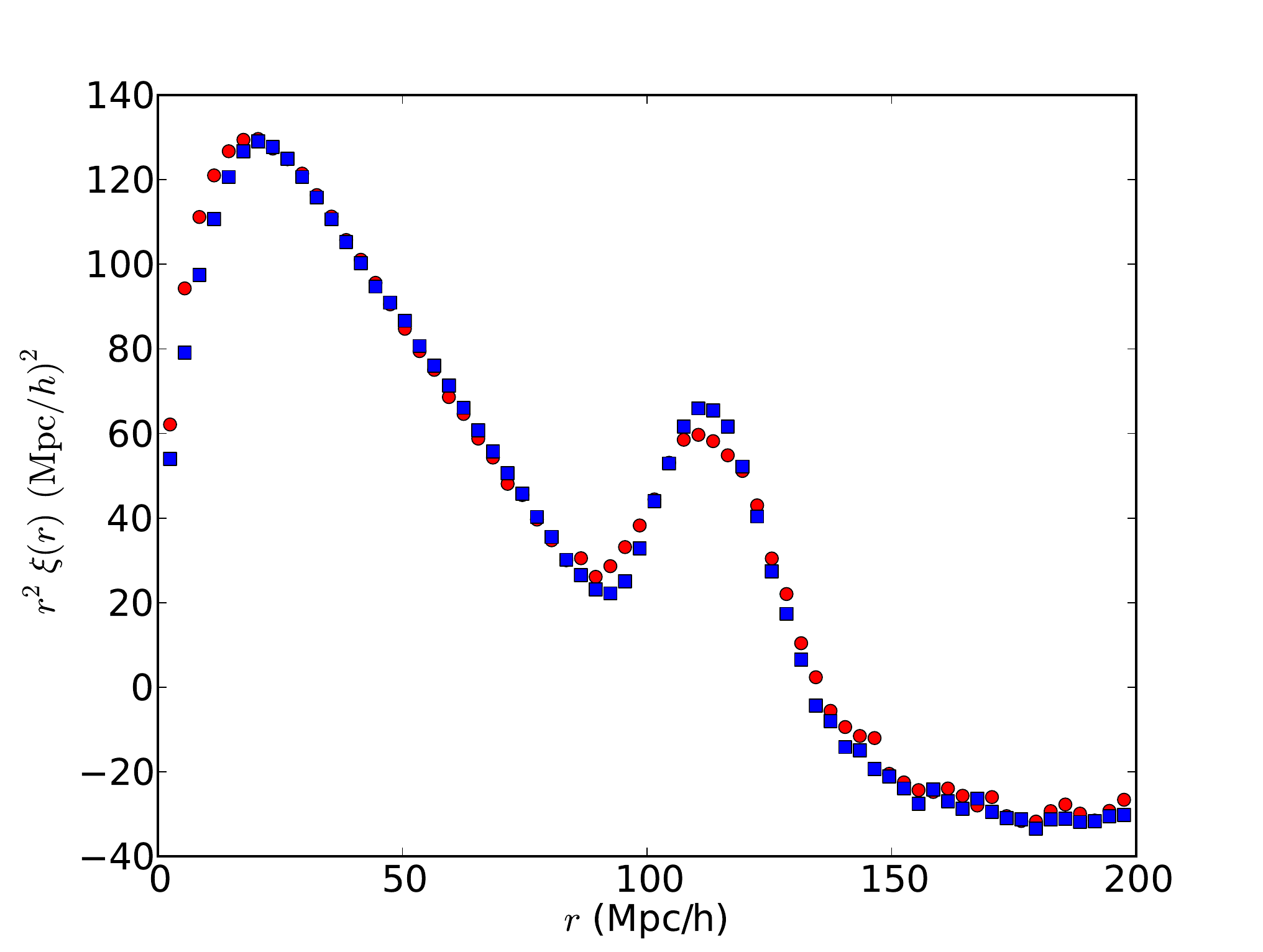}
  \includegraphics[width=3in]{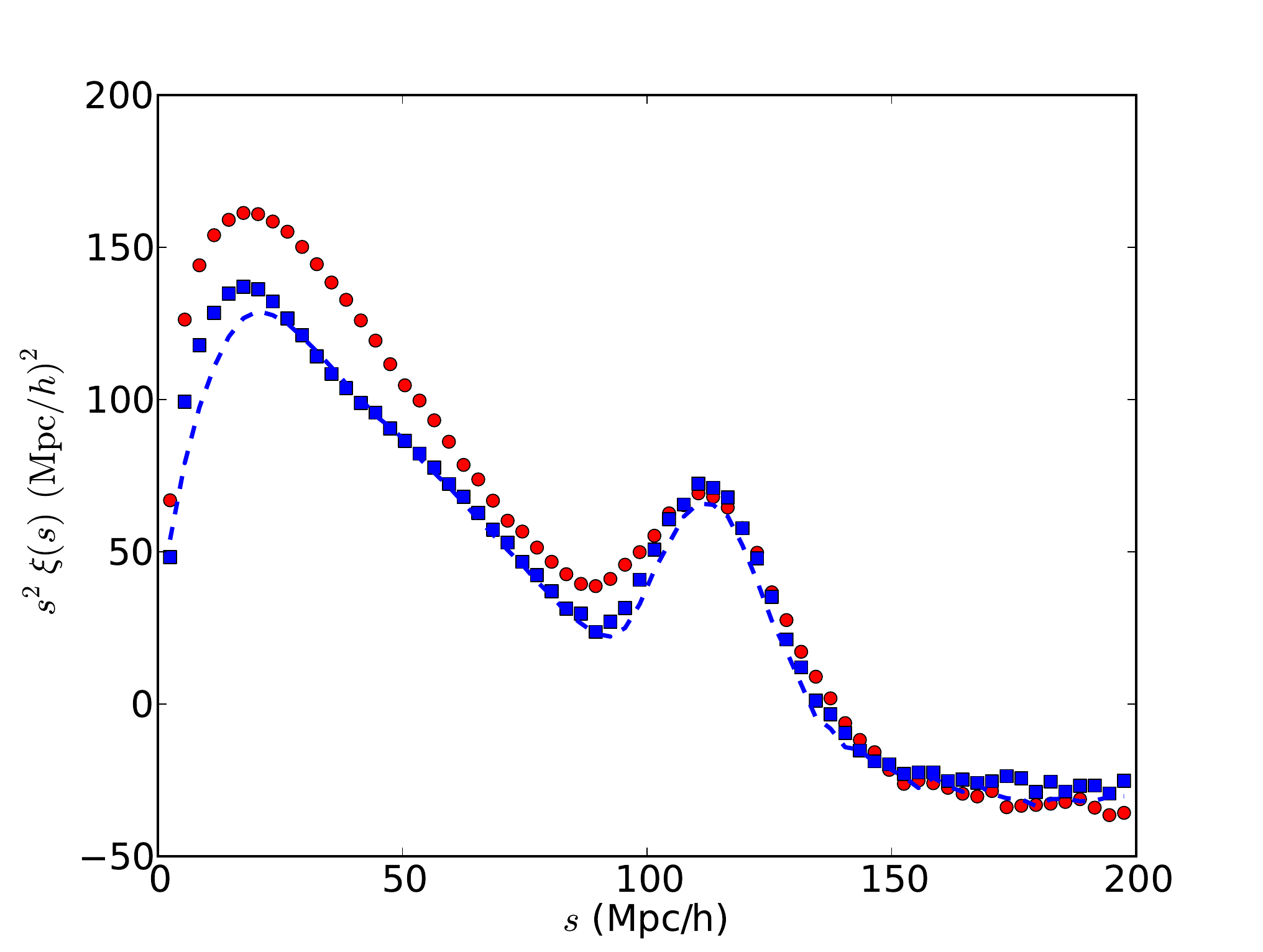}
  \caption{[left]The angle averaged correlation function in real space, before [red circles] and after [blue
  squares] reconstruction and averaging over the 160 LasDamas simulations. 
  The reconstruction algorithm assumes the default parameters described in the text. The acoustic feature
  is clearly sharpened after reconstruction. [right] Same as the left panel,
  except in redshift space. Also shown for comparison
  is the average reconstructed real-space correlation [dashed line]. In addition to sharpening the acoustic
  feature, the reconstruction algorithm also reduces the effects of redshift-space distortions on the correlation
  function.}
  \label{fig:plot_xi_lasdamas}
\end{figure*}

We compress the 2D correlation functions by averaging over angle; the resulting 
correlation functions (both in real and redshift space) are in Figure~\ref{fig:plot_xi_lasdamas}. In both cases, we 
observe the BAO feature sharpened after reconstruction. In the case of the redshift-space distortions, the overall
amplitude of the correlation function is also reduced due to the removal of linear redshift-space distortions. We
find that the reconstructed redshift-space correlation function does not match its real-space counterpart 
on small scales, indicating that the linear theory correction is breaking down on these scales. 
However, the agreement on large $(r > 30 {\rm Mpc}/h)$ scales is striking.

Figure~\ref{fig:plot_xi_lasdamas} has a useful interpretation as a redistribution of pairs of 
galaxies across
different scales. Recall that $r^2 \xi$ is proportional to the number of excess pairs (over a random distribution)
in an annulus of width $dr$ centered at $r$. Since reconstruction does not change the total number of pairs
but merely redistributes them over different scales, the area under these curves must be conserved. 
Comparing the correlation functions before and after reconstruction captures this redistribution of pairs. 
There are two 
effects worth noting, both of which are more easily noted in the real-space case. The first is a transferring
of pairs from small scales ($r < 20 {\rm Mpc}/h$) to intermediate scales ($r \sim\! 50 {\rm Mpc}/h$), apparent
in the fact that the unreconstructed correlation function is larger on small scales, with the trend reversed on 
intermediate scales. This is reconstruction reversing the infall of galaxies into overdensities. The second is 
that the unreconstructed correlation function is higher just before the BAO feature, due to pairs flowing 
out of the BAO feature. These flows are responsible for the smoothing of the BAO feature. The fact that 
the reconstructed correlation function is lower just before the BAO feature and then higher at the BAO peak
is from the fact that reconstruction has moved these objects back into the BAO ring.

One metric to quantify the degree of reconstruction is to compare the values of $\Sigma_{NL}$ (see Eq.~\ref{eqn:template})
before and after reconstruction. While $\Sigma_{NL}$ is poorly constrained in any single simulation, we 
can fit the average of the simulations before and after reconstruction. We find that $\Sigma_{NL}$ decreases by close to 
50\% from 8.1 to 4.4 Mpc/$h$. This improvement is in line with theoretical estimates \citep{2009PhRvD..79f3523P} and 
corresponds well with assumptions made for future surveys.

\begin{figure}
  \includegraphics[width=3in]{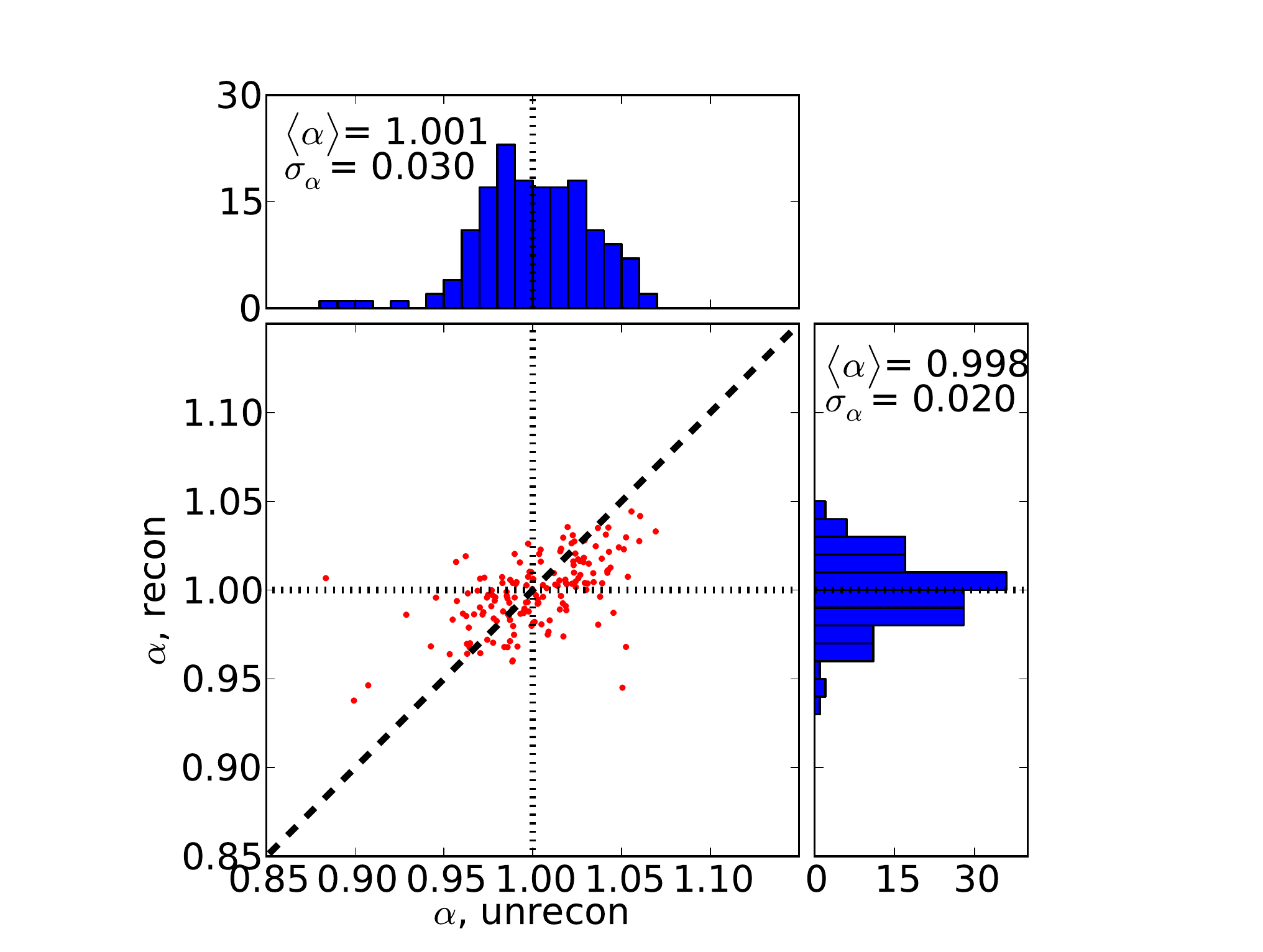}
  \caption{The distance scale, $\alpha$ estimated for the LasDamas simulations in real space, before (``unrecon'')
  and after (``recon'') reconstruction and their projected 1D distributions. The lines on the scatter
  plot mark the true ($\alpha=1$) values as well as the equality ($\alpha_{\rm recon} = \alpha_{\rm unrecon}$)
  line. We find no evidence of a statistical bias in the recovered $\alpha$ values; reconstruction
  reduces the scatter in these values.}
  \label{fig:lasdamas_alpha_real}
\end{figure}

\begin{figure}
  \includegraphics[width=3in]{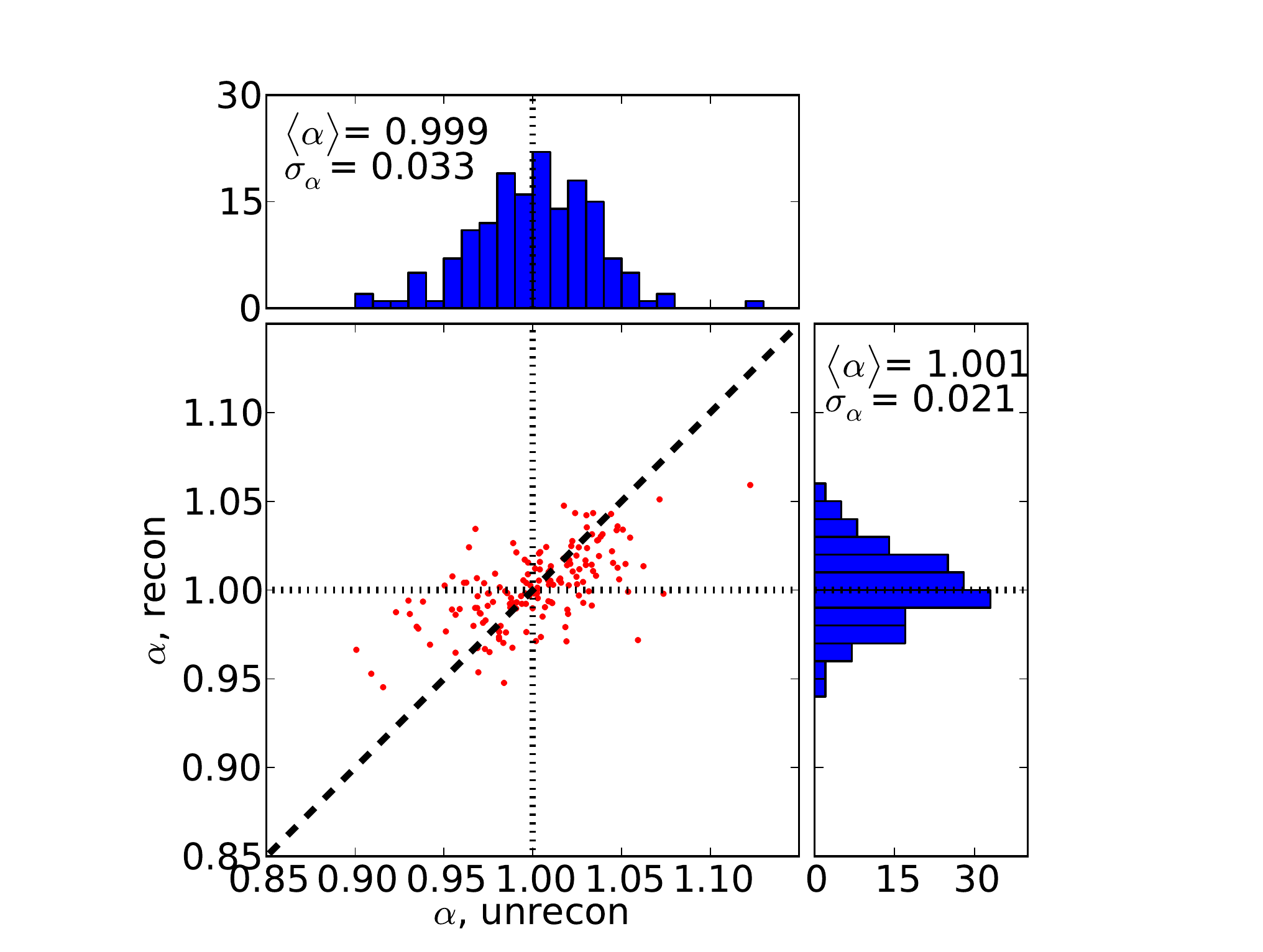}
  \caption{The same as Figure~\ref{fig:lasdamas_alpha_real} except now in redshift space. The scatter in the 
  recovered values is larger than the real-space case, with reconstruction still significantly reducing the
  scatter in these values.}
  \label{fig:lasdamas_alpha_red}
\end{figure}

Figures~\ref{fig:lasdamas_alpha_real} and \ref{fig:lasdamas_alpha_red} quantify the impact of reconstruction
on the inferred distance $\alpha$ in real and redshift space respectively. Recall that $\alpha$ is the 
estimated distance relative to a fiducial distance, which, in the
case of the simulations, we choose to be the true comoving distance to the
median redshift of the survey $z \sim\! 0.35$. Both the unreconstructed and reconstructed simulations 
yield unbiased distance estimates (i.e. $\langle \alpha \rangle = 1$) and the distances before 
and after reconstruction are clearly correlated with one another. Reconstruction, however, reduces the scatter
in $\alpha$ from 3.0\% to 2.0\% in real space and from 3.3\% to 2.1\% in redshift space, an improvement of 
between a factor of 1.5 to 1.7. These figures also demonstrate that reconstruction noticeably reduces the
number of outliers in the distance estimate, a direct effect of the increased significance of the BAO feature.

Before continuing, it is worth noting that the BAO scale in galaxies is expected to be biased at the 
$\sim\! 0.5\%$ level (or lower) \citep{2009PhRvD..80f3508P,2011ApJ...734...94M}. These biases are understood to arise from 
second-order terms in perturbation theory \citep{2008PhRvD..77b3533C, 
2009PhRvD..80f3508P}  and are expected to be reduced 
by reconstruction \citep{2009PhRvD..79f3523P,2009PhRvD..80l3501N,2011ApJ...734...94M}. 
However, the amplitude of these effects are much below what we can expect to observe
with a single SDSS sized dataset and are expected to be only marginally detectable with the full ensemble of simulations.
Table~\ref{tab:basic} verifies this expectation, demonstrating that any bias in the distance scale is less
than $0.2\%$, much below our statistical precision.
We therefore ignore these effects in all our subsequent analyses but note that they will become relevant 
as the statistical error decrease for future surveys.

\begin{figure}
  \includegraphics[width=3in]{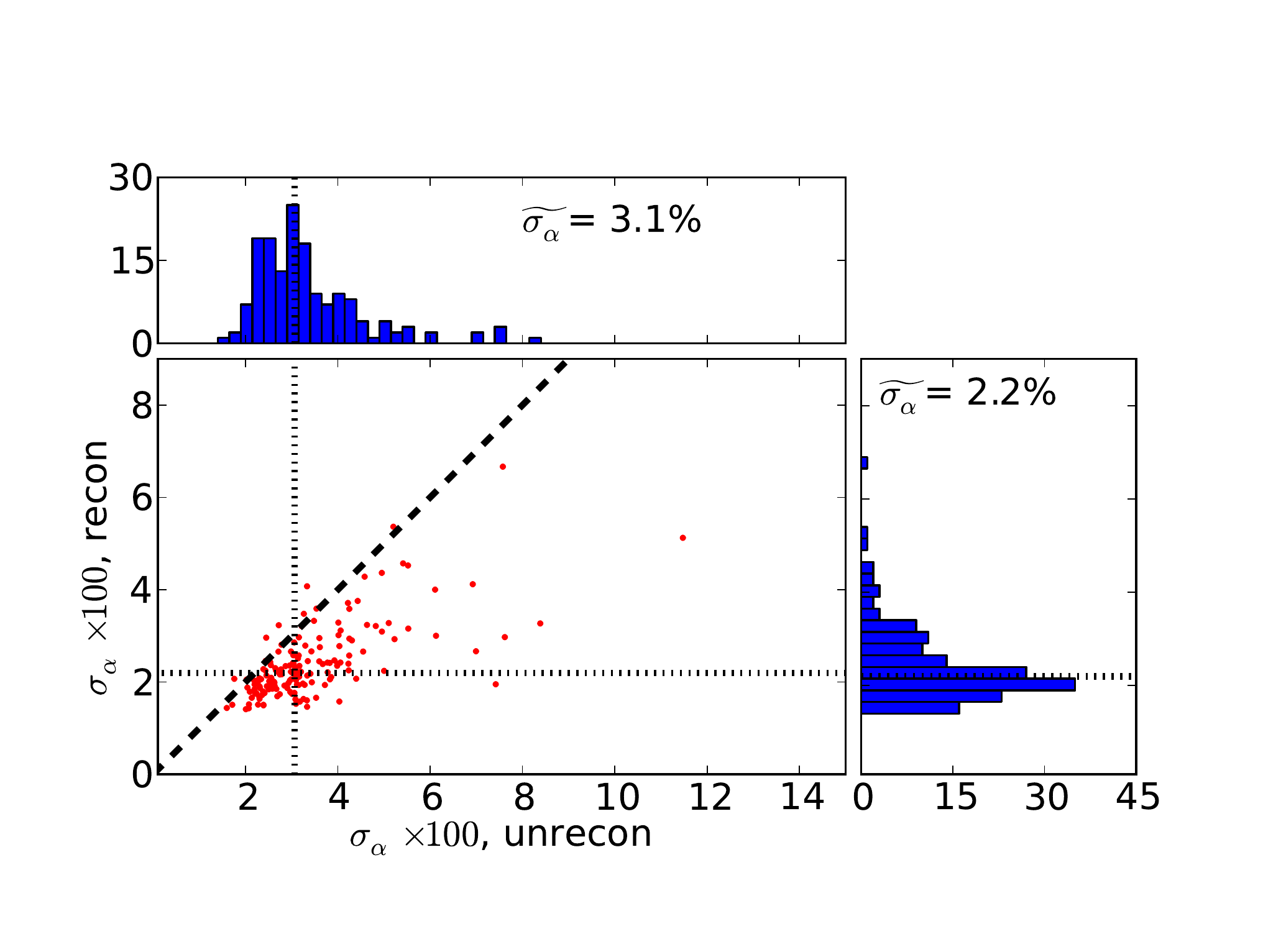}
  \caption{The error on the distance scale before and after reconstruction, 
  as individually estimated for each LasDamas simulation in real space. 
  The short dashed line has slope 1; the fact that 
  most points lie below the line demonstrates the efficacy of reconstruction. The 
  horizontal and vertical lines mark the median error before and after reconstruction.
  Note also the large
  scatter in the estimated errors, due to the still relatively low significance of the BAO detection
  in the DR7 survey volume.
  }
  \label{fig:lasdamas_sigma_alpha_real}
\end{figure}

\begin{figure}
  \includegraphics[width=3in]{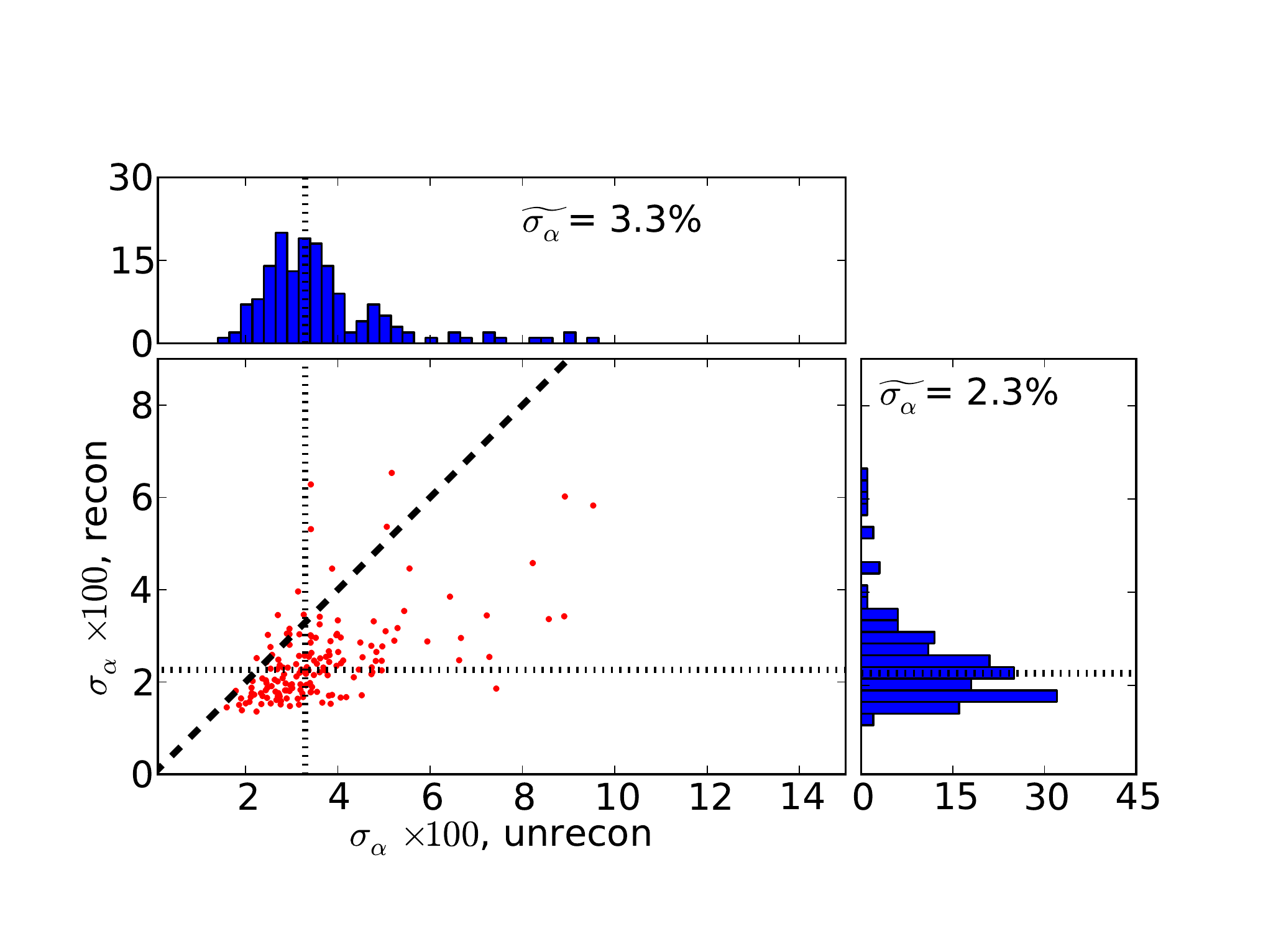}
  \caption{The same as Figure~\ref{fig:lasdamas_sigma_alpha_real} except in redshift space.}
  \label{fig:lasdamas_sigma_alpha_red}
\end{figure}

The above distance accuracies are ensemble averages. Given the still relatively low significance of the 
detection of a BAO feature, statistical fluctuations may make the BAO feature more or less prominent. Since
our fitting procedure marginalizes out smooth components in the correlation function, a less prominent 
BAO feature would result in a significantly degraded distance measurement. It is therefore interesting
to quantify the effect of reconstruction not just on the ensemble distance accuracies, but also on the 
distance accuracy estimated for each individual simulation. Figures~\ref{fig:lasdamas_sigma_alpha_real} and
\ref{fig:lasdamas_sigma_alpha_red} summarize this information for real and redshift space respectively. The 
median errors agree well with the errors estimated from the ensemble distance measurements, confirming the
validity of our error estimates. We also observe that, for the majority of the simulations ($\sim\!$ 98\%),
reconstruction reduces the distance error. For the remaining $\sim\!$ 2\%, we find that the errors are very 
similar to those obtained before reconstruction. Furthermore, a number of these are cases where the 
distance scale itself is poorly measured. Paper II discusses these cases in more detail. Table~\ref{tab:basic}
summarizes the above discussion, considering the recovered distances before and after reconstruction.

\begin{table}
  
\begin{tabular}{ccccc}
\hline
Type & $\langle \alpha-1 \rangle$  
     & $\sigma_{\alpha}$ 
     & $\widetilde{\sigma_{\alpha}}$ 
     & n($\sigma_{\alpha}$ $>$ 0.07) \\
\hline
Real, unrecon & \phantom{-}$0.1$ & \phantom{-}3.0 & \phantom{-}3.1 &   5 \\
Real, recon & -$0.2$ & \phantom{-}2.0 & \phantom{-}2.2 &   0 \\
Redshift, unrecon & -$0.1$ & \phantom{-}3.3 & \phantom{-}3.3 &   8 \\
Redshift, recon & \phantom{-}$0.1$ & \phantom{-}2.1 & \phantom{-}2.3 &   0 \\
\hline
\end{tabular}

  \caption{A summary of the effect of reconstruction on the distance estimates, in real and redshift space.
  The $\langle \alpha-1 \rangle$ and $\sigma_{\alpha}$ numbers are in percent.
  For all these cases, we reconstruct using our default choices of parameters.
  The second column is the bias in the distance in percent. The third column is the error in the distance
  estimated from the ensemble of the LasDamas simulations, while the fourth column is the median of the
  errors estimated per simulation, with the agreement between the two testing our error estimates. The last column
  shows the number of cases where the error in the distance is greater than $7\%$. In all of these metrics, 
  reconstruction improves the precision of the distance estimates.
  }
  \label{tab:basic}
\end{table}

\subsection{Robustness to Reconstruction Parameters}

All of the above has assumed our default choices of parameters for reconstruction - a smoothing scale of 15 Mpc/$h$, 
the measured value of the galaxy bias ($b=2.2$) and the input value of $f$. We also assume a concordance cosmology power spectrum
in order to generate the constrained Gaussian realization. We explore the effects of varying these below. We use the simulations
to determine the optimal smoothing scale and demonstrate that reconstruction is robust to variations in the other parameters. 

The most important of the reconstruction parameters is the scale used to smooth the density field 
before estimating the potential. For scales too small, the fidelity of the potential reconstruction
is affected by noise from the finite numbers of galaxies. At the other extreme, 
over-smoothing reduces the estimated displacements (an infinite smoothing scale leaves the galaxies at their
original positions) and reduces the effectiveness of reconstruction. In the limit of a high number
density of galaxies, the above argument suggests choosing a smoothing scale as small as possible; however, 
as argued in \citet{2007ApJ...664..660E}, the bulk of the smoothing comes from scales between $k=0.02$ to $0.2$ $h$/Mpc, 
suggesting that one reaches diminishing returns for scales smaller than $\sim\! 5$ Mpc/$h$. The number 
density of the LRG sample is $\le 10^{-4} h^{3}/{\rm Mpc}^3$, which implies that the shot noise power spectrum 
crosses at a scale of $\sim\! 0.15 h/{\rm Mpc}$, suggesting smoothing on scales larger than $\sim\! 10 {\rm Mpc}/h$.

\begin{table}
  
\begin{tabular}{ccccc}
\hline
Smoothing & $\langle \alpha-1 \rangle$  
     & $\sigma_{\alpha}$ 
     & $\widetilde{\sigma_{\alpha}}$ 
     & n($\sigma_{\alpha}$ $>$ 0.07) \\
\hline
15 Mpc/h & \phantom{-}$0.1$ & \phantom{-}2.1 & \phantom{-}2.3 &   0 \\
20 Mpc/h & \phantom{-}$0.4$ & \phantom{-}2.3 & \phantom{-}2.5 &   0 \\
25 Mpc/h & \phantom{-}$0.6$ & \phantom{-}2.6 & \phantom{-}2.6 &   0 \\
\hline
\end{tabular}

  \caption{Analogous to Table~\ref{tab:basic}, except as a function of the smoothing 
  scale input to the redshift-space reconstruction. As before, 
  the $\langle \alpha-1 \rangle$ and $\sigma_{\alpha}$ numbers are in percent. In all cases, the
  distance is unbiased, but as expected, we find the error increasing as a function of the smoothing
  scale. Reconstruction after smoothing at 10 Mpc/$h$ distorts the shape of the correlation 
  function and our model is no longer a good fit to the shape. We use 15 Mpc/$h$ as our 
  fiducial smoothing scale.
  }
  \label{tab:smooth}
\end{table}

\begin{figure*}
  \includegraphics[width=3.0in]{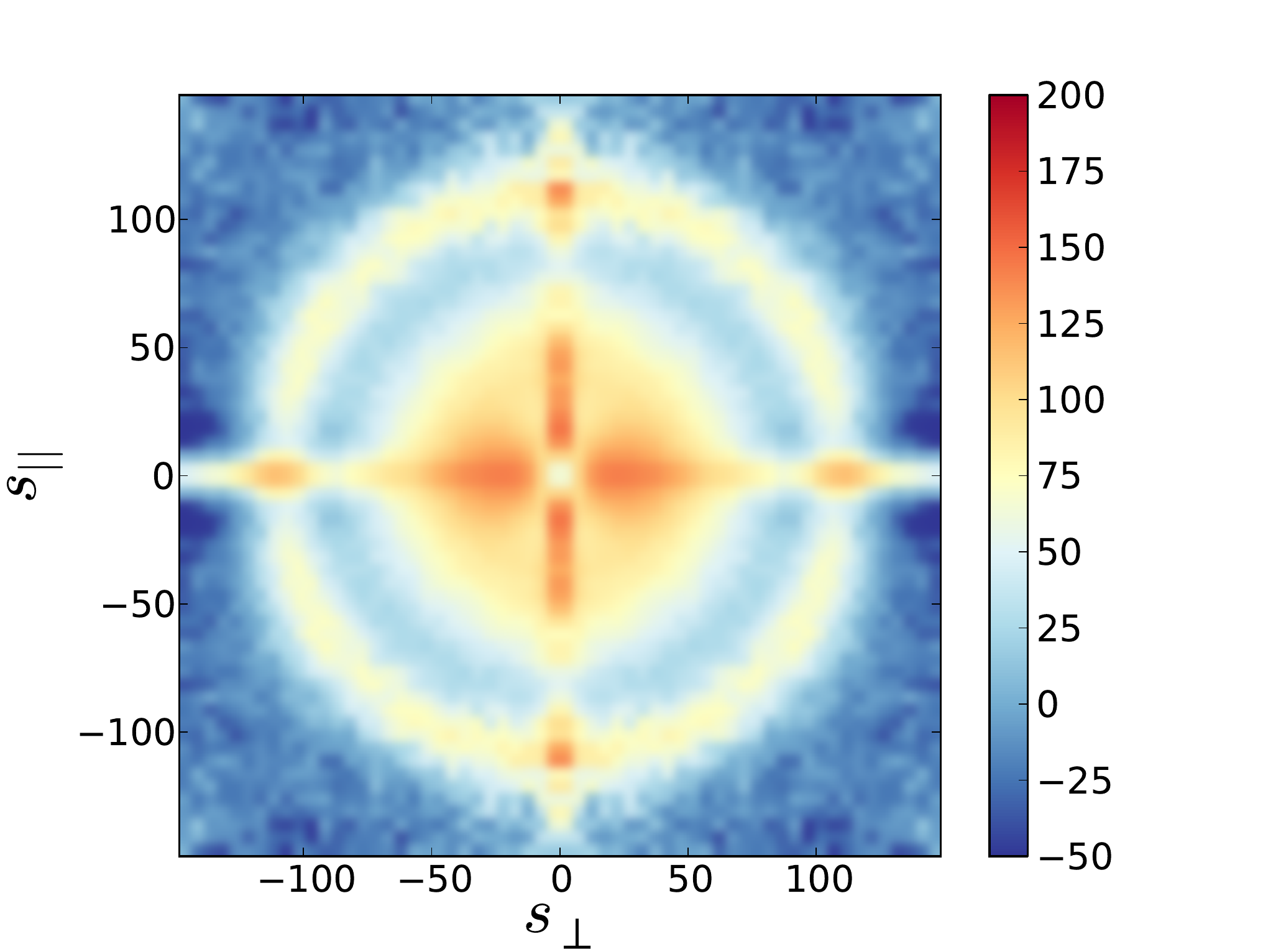}
  \includegraphics[width=3.0in]{plots/lasdamas_wp_recon_s15}
  \includegraphics[width=3.0in]{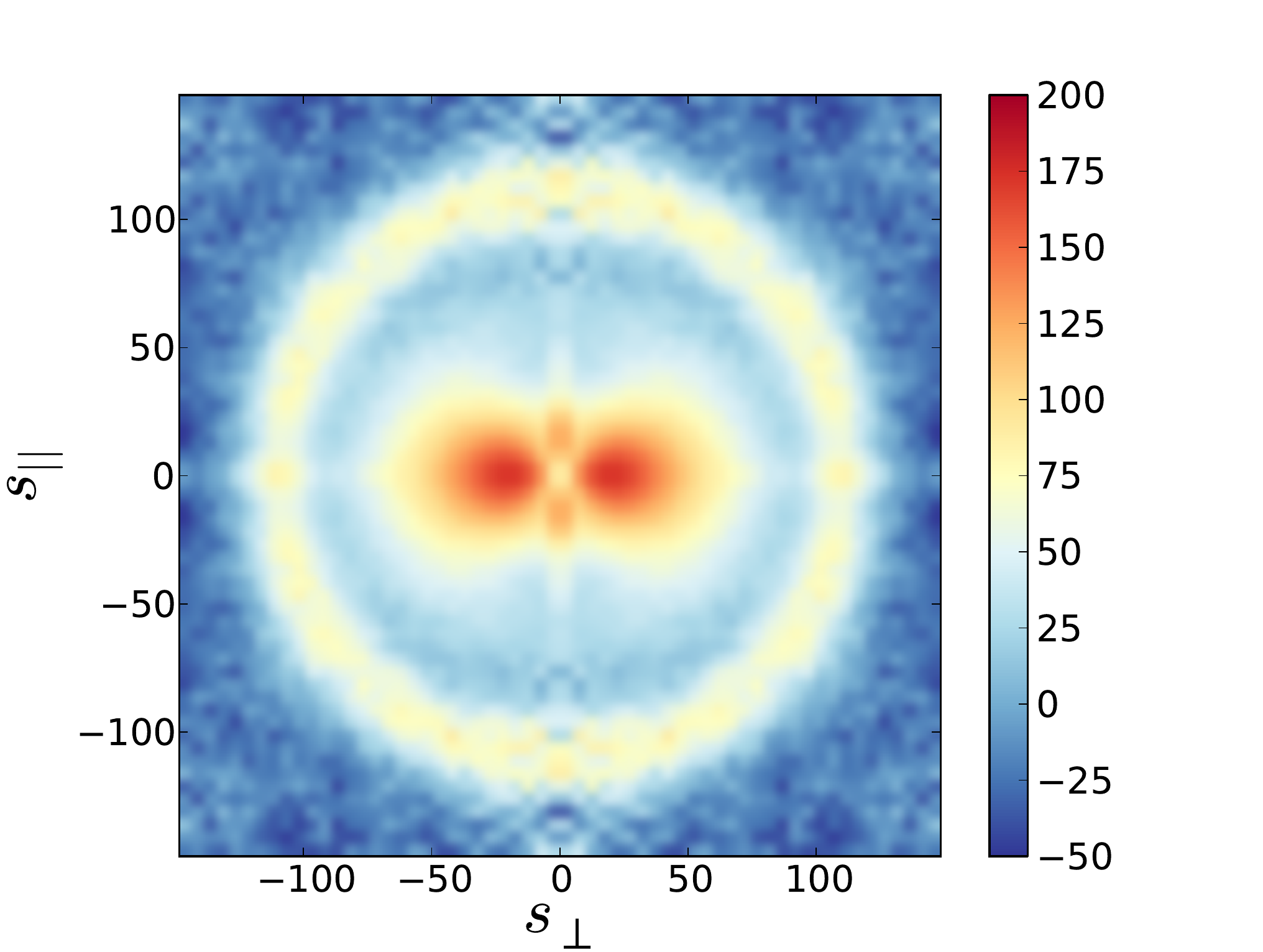}
  \includegraphics[width=3.0in]{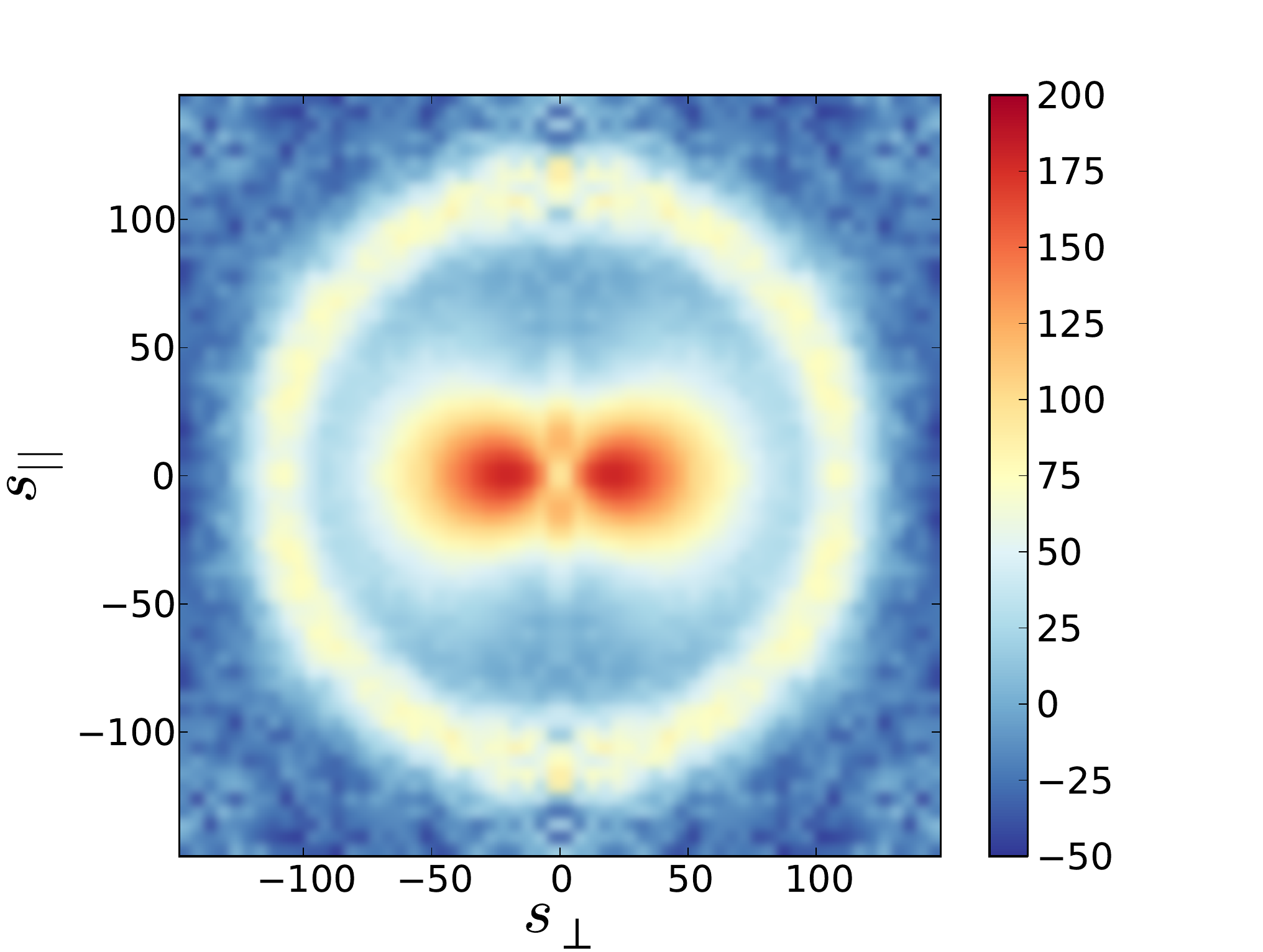}
  \caption{The impact of changing the smoothing length on reconstruction for the 2D correlation functions. 
  The panels correspond to smoothing scales of 10 (top left), 15 (top right), 20 (bottom left) and 25 (bottom right) Mpc/$h$.
  Note the prominent distortions in the correlation function at small scales for the 
  10 Mpc/$h$ smoothing scale and the degraded reconstruction for the 25 Mpc/$h$ case. Our fiducial choice is the 15 Mpc/$h$ case.}
  \label{fig:wp_lasdamas_smooth}
\end{figure*}

Figure~\ref{fig:wp_lasdamas_smooth} plots the reconstructed 2D correlation functions for different choices of the smoothing
scale. For large smoothing scales, the degree of reconstruction is clearly degraded, although even in these cases, the BAO
feature is still enhanced relative to the case of no reconstruction. This emphasizes the fact that it is large-scale 
flows that are responsible for the erasure of the BAO feature; even a large smoothing scale can effectively reverse these. 
At the other extreme at 10 Mpc/$h$, we find that reconstruction can strongly distort the correlation function on small scales.
A useful picture to understand the distortions at small perpendicular separation is to remember that reconstruction 
effectively repels close pairs of particles. Since the smoothing scale averages the shot noise in the input density field, 
insufficiently smoothing the field pushes apart noise fluctuations with an additional enhancement in the line of sight direction
(generating its own Fingers of God). 

The second feature is the excess at $r_{||} = 0$. A examination of Figure~\ref{fig:wpav} shows that traces of this feature exist
in the real-space correlation function as well. Unlike the redshift-space case, the reconstruction procedure in real space does
not have a preferred direction. This feature can be traced back to our definition of the radial selection function. Since we 
do not know the mean density of the galaxies as a function of redshift, we simply define it by randomly resampling the observed 
galaxy redshifts. This suppresses radial density fluctuations and can create features in the transverse direction. As before, 
increasing the smoothing scale reduces both of these.

\begin{figure}
  \includegraphics[width=3in]{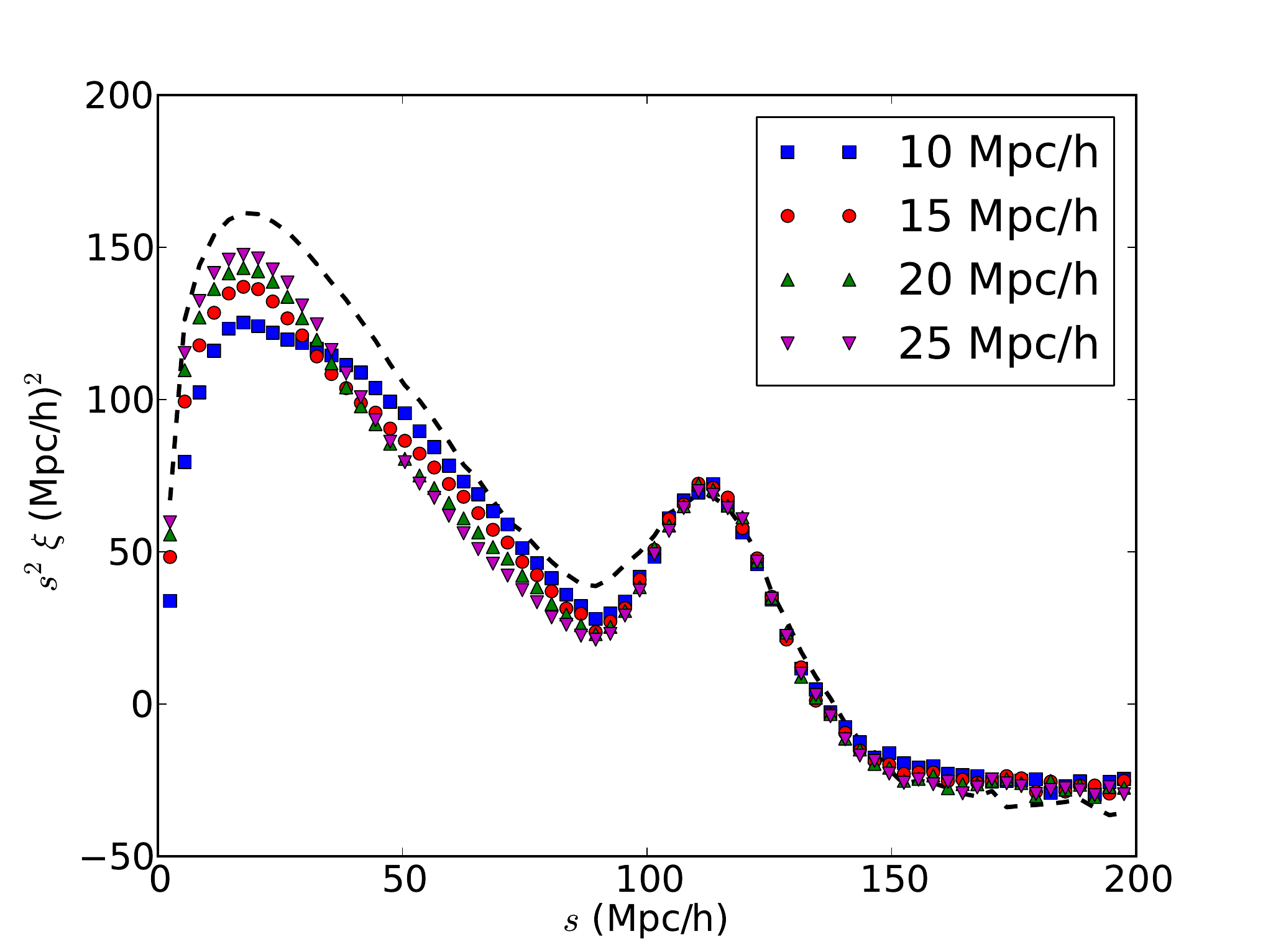}
  \caption{The averaged reconstructed correlation functions in redshift space for the LasDamas
  simulations, as a function of the reconstruction smoothing scale. Also plotted for reference [dashed line]
  is the unreconstructed correlation function. Our fiducial smoothing length is 15 Mpc/$h$.} 
  \label{fig:xi_lasdamas_smooth}
\end{figure}

Figure~\ref{fig:xi_lasdamas_smooth} plots the angle averaged correlation function. We find that the correlation function 
is strongly distorted on small scales for the 10 Mpc/$h$ smoothing scale; increasing the smoothing scales removes these 
distortions. Given our desire to choose the smallest smoothing scale possible, we adopt 15 Mpc/$h$ as our fiducial choice. 
Table~\ref{tab:smooth} summarizes the distance constraints as a function of smoothing scale and shows the same trends 
discussed above.

While the smoothing scale is the most important parameter input to the reconstruction algorithm, there are
a number of other inputs - the galaxy bias $b$, the logarithmic growth rate of structure $f$, and the power spectrum 
used for the constrained realizations. The galaxy bias and the input power spectrum are constrained by measurements
of the unreconstructed correlation function, while $f \sim\! \Omega_{M}^{0.55}$ can be constrained by CMB measurements. 
While we adopt these fiducial values for our measurements, an immediate question is how sensitive reconstruction is 
to these choices. Table~\ref{tab:params} summarizes the impact of changing these values. We vary the bias and $f$ by 
$\pm 20\%$ and we consider two different choices for a fiducial power spectrum - a power spectrum with the BAO 
features erased \citep{1998ApJ...496..605E} and a constant $P(k) = 10^{4} ({\rm Mpc}/h)^3$. 
Of these, the bias has the strongest effect on reconstruction. An overestimate of the bias results in an underestimate 
of the density field, reducing the derived displacement field. On the other hand, underestimates of the bias result in 
an overestimate of the density field and over-corrects the displacements. 
While the distance scale obtained is still unbiased, Table~\ref{tab:params} demonstrates 
including the uncertainty in the galaxy bias in reconstruction would increase the statistical uncertainty of the 
measurement. However, as Table~\ref{tab:params} also shows, even for large variations in the bias (recall that 
a 20\% uncertainty in the bias corresponds to a 40\% uncertainty in the normalization of the correlation function, 
much larger than most estimates of the bias), this additional scatter is subdominant to our errors. We also 
observe the impact of mis-estimating the bias on the inferred error in the mocks is much less than our 
quoted statistical errors.

\begin{figure}
  \includegraphics[width=3in]{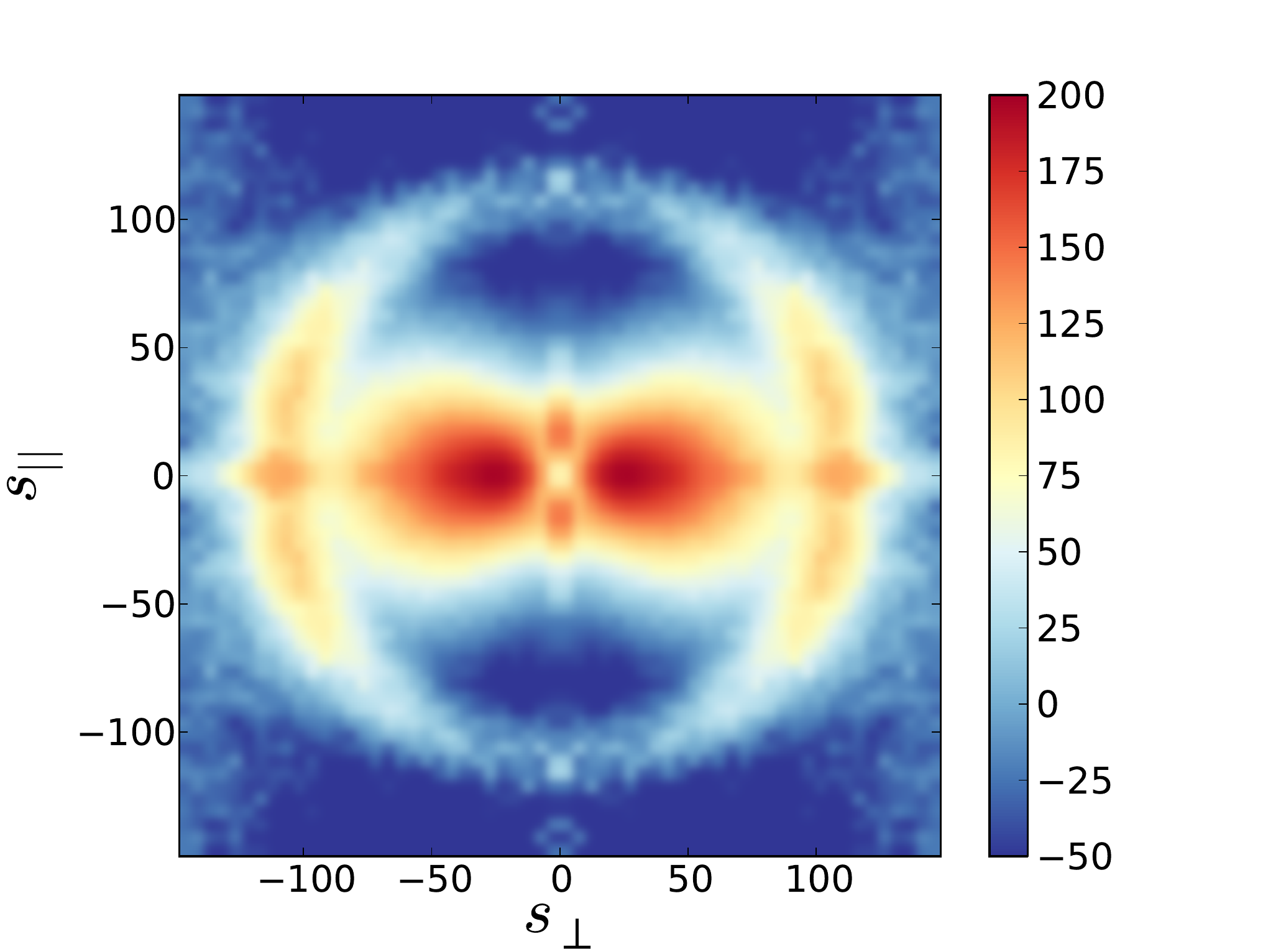}
  \caption{The 2D reconstructed correlation function, but without the redshift-space distortion corrections (i.e. setting $f=0$
  in the reconstruction algorithm). While the BAO feature is more prominent, redshift-space distortion still strongly distorts the 
  BAO feature.}
  \label{fig:recon_fval00}
\end{figure}

One of the most prominent effects of reconstruction was the restoration of the isotropy of the correlation function. 
Since this depends on the choice of $f$, an immediate question is how sensitive reconstruction is to the particular 
choice. Figure~\ref{fig:recon_fval00} plots the 2D correlation function with $f=0$. This turns off both the 
modifications to the continuity equation in redshift space, as well as the additional line of sight displacements
that correct for the redshift-space distortions.
We find that the BAO feature is still improved, although the correlation function is still strongly distorted from 
isotropy. Less drastic variations in the value of $f$ (although larger than our current errors on $\Omega_{m}$) are
in Table~\ref{tab:params}. Even more strongly than in the case of the bias, we find that this additional error is 
much smaller than our statistical error, demonstrating the robustness of reconstruction.

Finally, we consider varying the inputs to the constrained realizations. We consider two alternative power spectra - 
one with the BAO feature erased and the other with no clustering signal - and we find that reconstruction is robust 
to these choices as well. These results are dependent on the geometry of the survey and could possibly change for different geometries.

\begin{table}
  
\begin{tabular}{lcccc}
\hline
Case  
     & $\sigma(\Delta \alpha)$
     & $r$ 
     & $\sigma(\Delta \sigma_{\alpha}) $ 
     & $r$ \\
\hline
b=1.8 (-20\%) & \phantom{-}0.9 & \phantom{-}0.92 & \phantom{-}0.3 & \phantom{-}0.94 \\
b=2.6 (+20\%) & \phantom{-}0.6 & \phantom{-}0.97 & \phantom{-}0.2 & \phantom{-}0.97 \\
f=0.5 (-20\%) & \phantom{-}0.2 & \phantom{-}1.00 & \phantom{-}0.1 & \phantom{-}1.00 \\
f=0.8 (+20\%) & \phantom{-}0.2 & \phantom{-}1.00 & \phantom{-}0.1 & \phantom{-}1.00 \\
Pk, no-wiggle & \phantom{-}0.0 & \phantom{-}1.00 & \phantom{-}0.0 & \phantom{-}1.00 \\
Pk, shot-noise & \phantom{-}0.3 & \phantom{-}0.99 & \phantom{-}0.1 & \phantom{-}0.99 \\
\hline
\end{tabular}

  \caption{The impact of changing the values of the galaxy bias, $f$ and the input power spectrum for the constrained Gaussian
  from their fiducial values. For each mock, we compute the difference in distance, $\Delta \alpha$ and error, $\Delta \sigma_{\alpha}$ 
  from the fiducial choice of parameters. We report both the scatter in these quantities (in percent) as well as the cross
  correlation $r$ with the fiducial case. In all cases, we see that changing these parameters yields distances and errors
  that are almost perfectly correlated with those obtained using our fiducial parameters. The largest effect is seen for the galaxy
  bias which directly controls the displacement field.}
  \label{tab:params}
\end{table}

\section{Reconstructing Data}
\label{sec:dr7}

\begin{figure*}
  \includegraphics[width=3in]{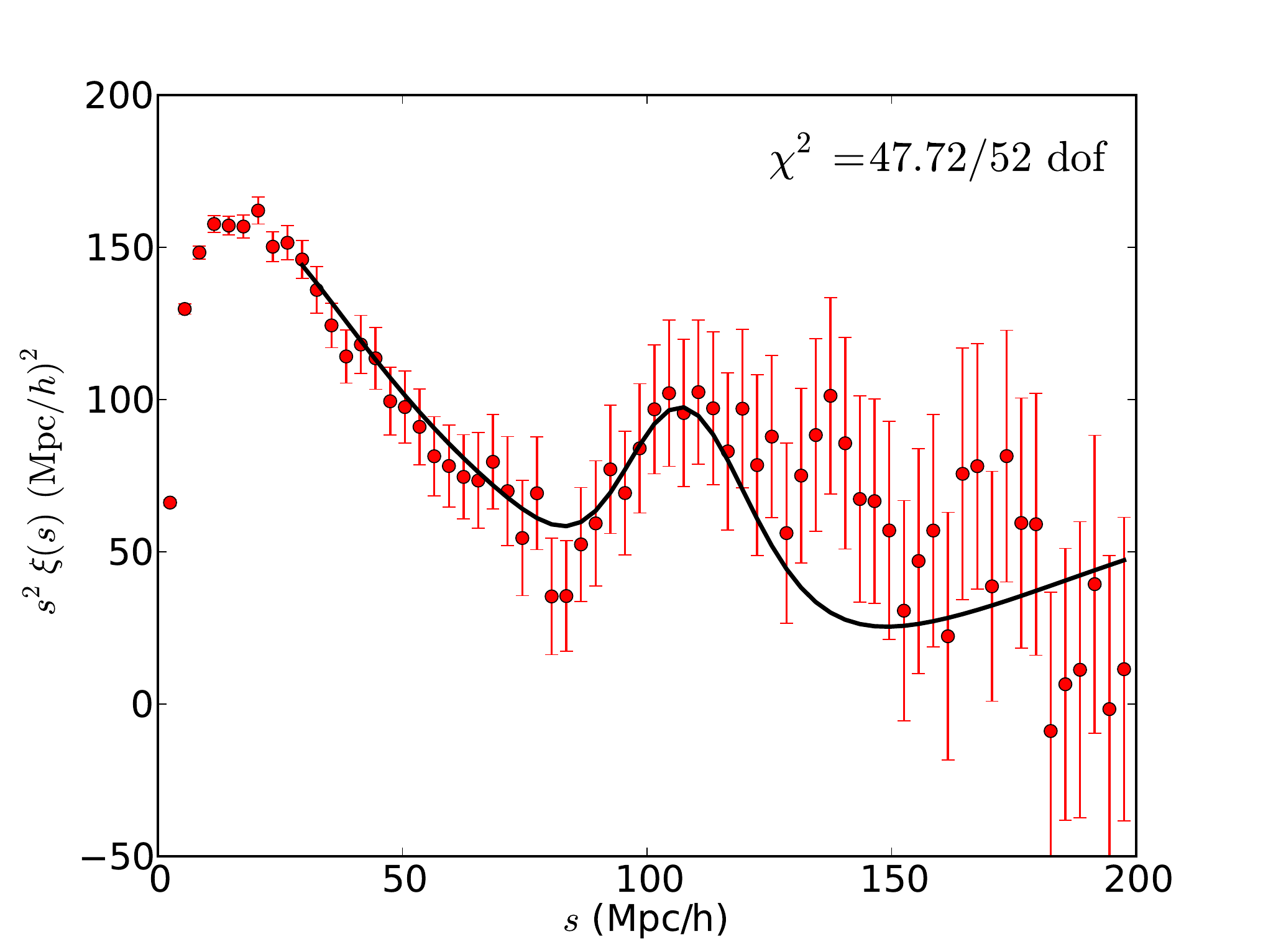}
  \includegraphics[width=3in]{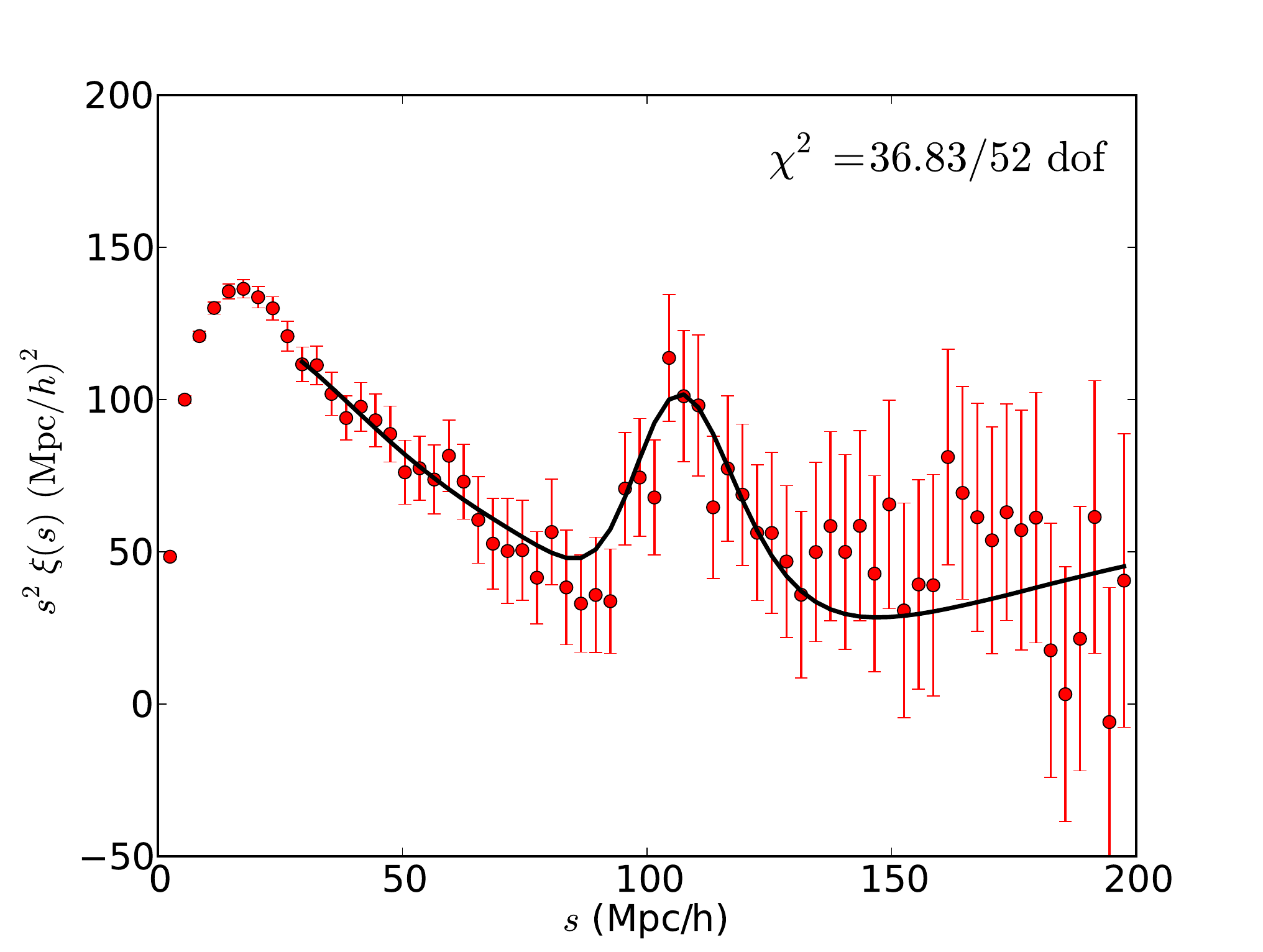}
  \caption{The unreconstructed [left] and reconstructed [right] DR7 angle averaged correlation function. 
  The error bars are the standard deviation of the 160 LasDamas simulations. These errors
  are however highly correlated from bin to bin and therefore no conclusions as to 
  significance should be drawn from these figures. The solid line is the best 
  fit model to these data. As in the 
  simulations, the acoustic feature appears sharpened.
  }
  \label{fig:dr7_xi}
\end{figure*}

We now apply reconstruction to the DR7 data set. Our fiducial choice of parameters is a 15 Mpc/$h$ smoothing length, 
a galaxy bias of 2.2 and the WMAP7 cosmology. Figure~\ref{fig:dr7_xi} plots the angle averaged DR7 correlation function 
before and after reconstruction. We find that reconstruction on the DR7
data demonstrates the same features seen in the LasDamas simulations. The amplitude of the intermediate-scale
correlation function decreases due to the correction of redshift-space distortions, while the transition into the 
BAO feature at $\sim\! 80-100$ Mpc/$h$ is sharpened. 

\begin{figure*}
  \includegraphics[width=3in]{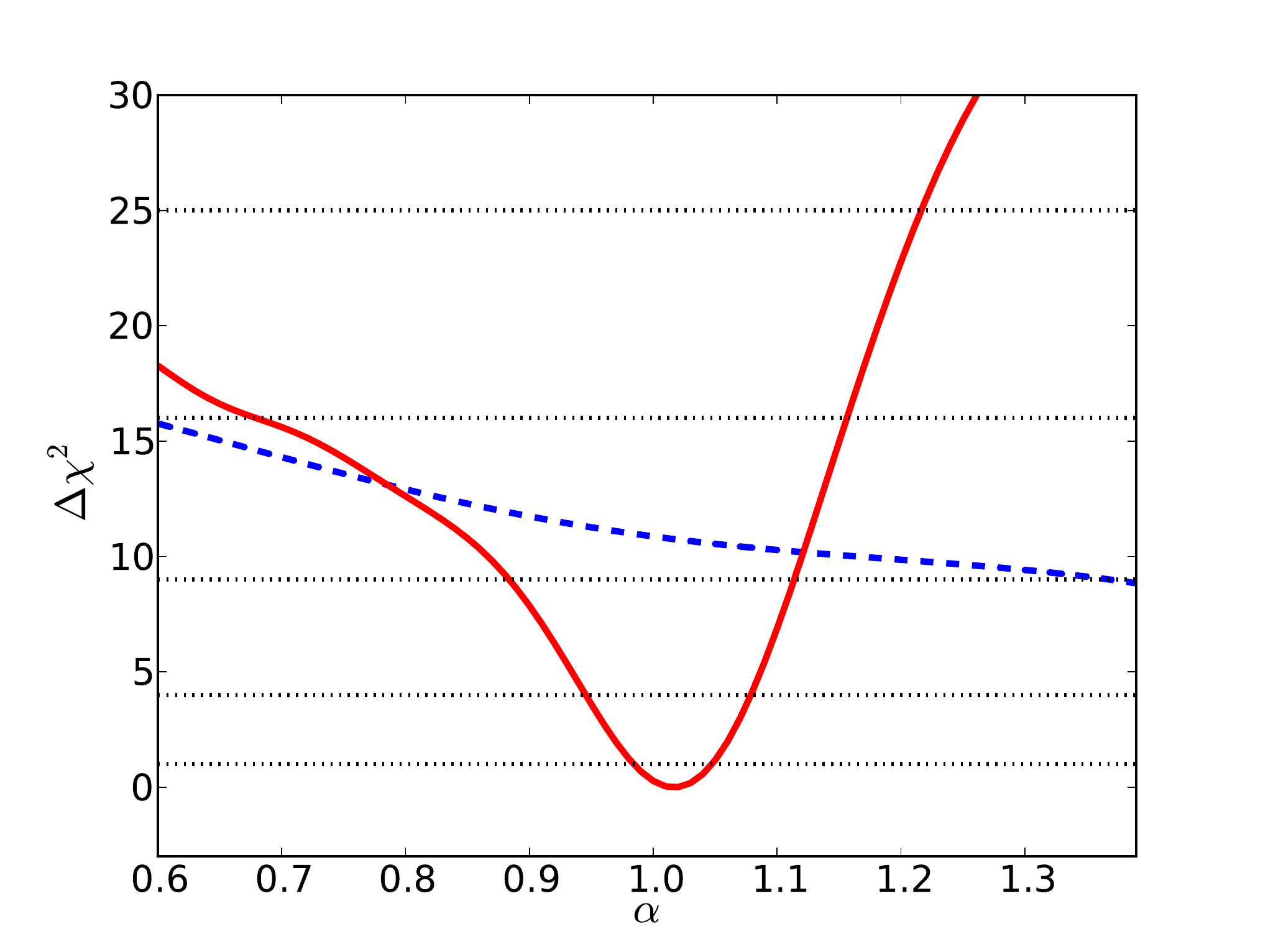}
  \includegraphics[width=3in]{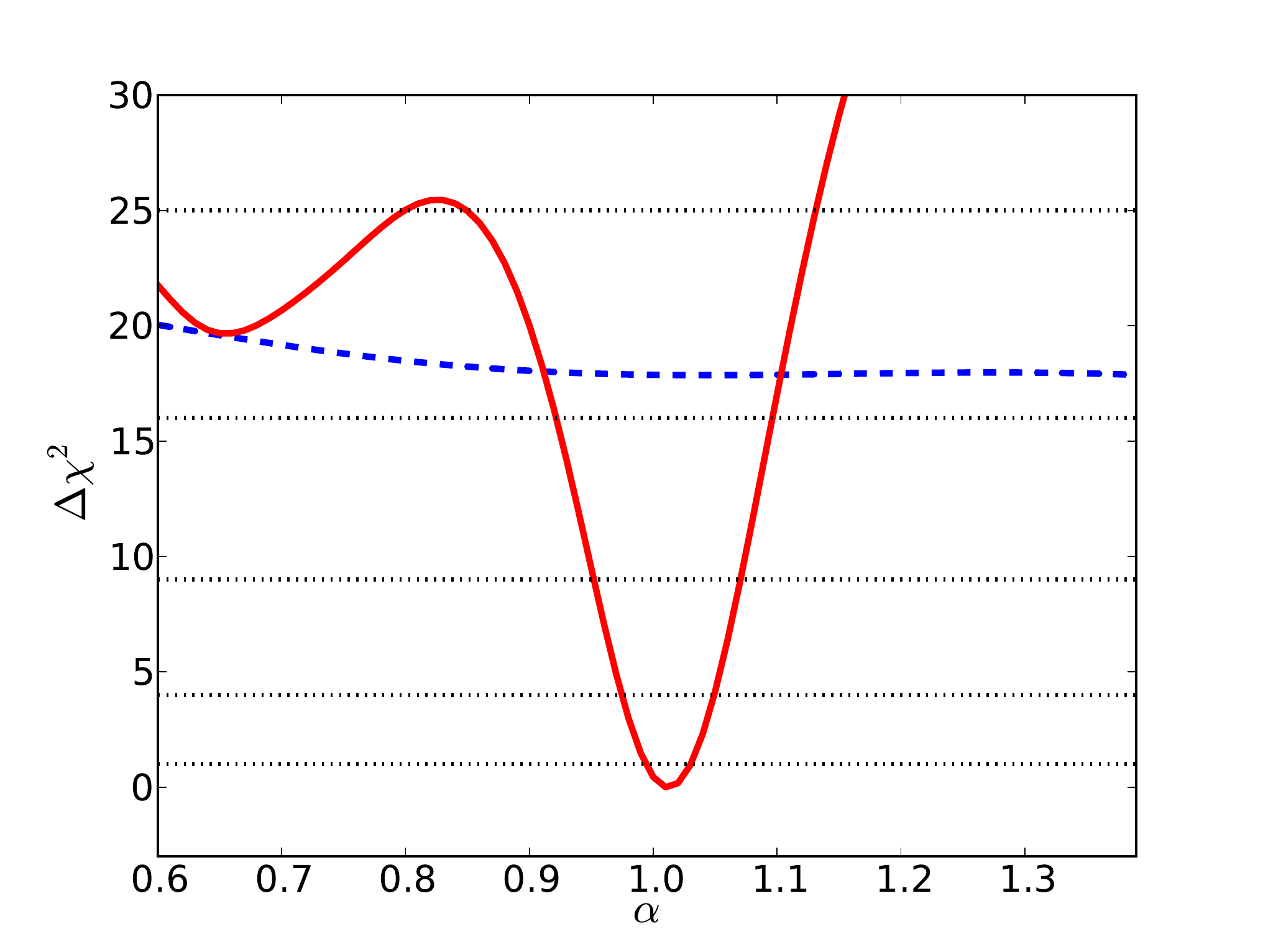}
  \caption{The $\Delta \chi^2$ surface as a function of $\alpha$ before [left] and after [right] reconstruction. 
  The solid [red] line uses our default template with a BAO feature in the correlation function, while the dashed [blue]
  line uses the ``no-wiggle'' form of \citet{1998ApJ...496..605E}. Reconstruction narrows the $\chi^2$ minimum, reflecting the 
  improved distance constraints. The difference between the solid and dashed lines estimates the significance 
  of the detection of the BAO feature; the horizontal dotted lines mark the 1 to 5 $\sigma$ significance level. Reconstruction
  increases the detection significance from 3.3 to 4.2 $\sigma$. }
  \label{fig:dr7_sig}
\end{figure*}

The correlated nature of the errors makes it difficult to quantitatively assess the impact of reconstruction on these
data. Figure~\ref{fig:dr7_sig} plots the $\chi^2$ surface for $\alpha$ both before and after reconstruction. We note that 
the $\chi^2$ minimum after reconstruction is visibly narrower, indicating an improvement in the distance constraints. 
This improvement is also summarized in the first two lines of Table~\ref{tab:dr7_basic} which shows that reconstruction 
reduces the distance error from 3.5\% to 1.9\%. These distance constraints are also consistent with the errors 
estimated from the LasDamas simulations.

Figure~\ref{fig:dr7_sig} also plots the $\chi^2$ surface for a template without a BAO feature, using the ``no-wiggle''
form of \citet{1998ApJ...496..605E}. The lack of a well defined minimum either before or after reconstruction 
indicates that our distance constraints are indeed coming from the presence of a BAO feature and not any broad 
band features in the correlation function. The difference in $\chi^2$ between the templates with and without 
a BAO feature also provides an estimate of the significance of the BAO detection in these data. Reconstruction 
improves this detection significance from 3.3$\sigma$ (consistent with previous measurements) to 4.2$\sigma$. This 
is not the only measure of the detection significance possible; Paper II discusses these in more detail.

\begin{table*}
  
\begin{tabular}{lccc}
\hline
Case  
     & $\alpha-1 (\times 100)$ & $D_{V}/r_{s}$  & $D_{V}$ (Gpc) \\
\hline
Unrecon & $1.3\pm3.5 $ & $ 8.89\pm0.31 $  & $1.358 \pm 0.047$ \\
\bf{Recon} &  $\mathbf{1.2\pm1.9}$ & $\mathbf{8.88\pm0.17}$ &$\mathbf{ 1.356 \pm 0.025}$ \\
\hline
Smoothing, 20 Mpc/h & $0.9\pm2.1 $ & $ 8.85\pm0.18 $  & $1.352 \pm 0.028$ \\
b=1.8 (-20\%) & $1.4\pm2.0 $ & $ 8.89\pm0.18 $  & $1.358 \pm 0.027$ \\
b=2.6 (+20\%) & $1.4\pm1.9 $ & $ 8.89\pm0.16 $  & $1.359 \pm 0.025$ \\
f=0.5 (-20\%) & $1.1\pm1.9 $ & $ 8.87\pm0.16 $  & $1.355 \pm 0.025$ \\
f=0.8 (+20\%) & $1.5\pm1.9 $ & $ 8.90\pm0.16 $  & $1.360 \pm 0.025$ \\
Pk, no-wiggle & $1.2\pm1.9 $ & $ 8.88\pm0.17 $  & $1.356 \pm 0.025$ \\
Pk, shot-noise & $1.5\pm1.9 $ & $ 8.90\pm0.17 $  & $1.360 \pm 0.026$ \\
\hline
$\Omega_{M}$ = 0.20 & $15.9\pm2.4 $ & $ 8.93\pm0.18 $  & $1.377 \pm 0.028$ \\
$\Omega_{M}$ = 0.35 & $-7.6\pm1.8 $ & $ 8.93\pm0.17 $  & $1.378 \pm 0.026$ \\
\hline
\end{tabular}

  \caption{The comoving distance to $z=0.35$, expressed (i) as $\alpha \equiv D_{V}/r_{s}/(D_{V}/r_{s})_{\rm fid}$
  (ii) $D_{V}/r_{s}$ and (iii) $D_{V}$ assuming the fiducial value of $r_{s}$ (and ignoring errors). The 
  first group of numbers compares the unreconstructed and reconstructed cases (assuming our default 
  reconstruction parameters). The distances obtained are consistent, but reconstruction reduces the 
  error by a factor of 1.8, resulting in a distance precise to 1.9\%. The second group explores the 
  impact of changing the various reconstruction parameters (as in Tables~\ref{tab:smooth} and ~\ref{tab:params};
  both the distance and its error are robust to any of these changes. Finally, we consider changing 
  the fiducial cosmology (keeping the physical densities fixed). In this case, we expect $\alpha$ to change, 
  since our fiducial distance changes, but the derived distances (both absolute and relative to the 
  sound horizon are unchanged. Note that the errors do change, reflecting changes in the volume 
  relative to the acoustic scale.}
  \label{tab:dr7_basic}
\end{table*}

As before, we would like to demonstrate the robustness of the results to the various parameters of the reconstruction
algorithm. Table~\ref{tab:dr7_basic} lists the recovered distances varying the smoothing scale, input bias, growth rate ($f$), 
and  prior power spectrum; for each of these cases, we recover distances consistent with the fiducial choices of 
parameters. 

Our final test is the impact of the assumed fiducial cosmology. We consider two cases in Table~\ref{tab:dr7_basic}: 
flat $\Lambda$CDM cosmologies with $\Omega_{M} = 0.2$ and $0.35$. In both of these cases, we adjust the Hubble 
constant and the baryon density $\Omega_{b}$ to keep the physical densities $\Omega_{b} h^2$ and $\Omega_{M} h^2$ 
equal to their WMAP7 values. This prescription leaves the CMB unchanged, but alters the distance-redshift relation. 
We find that the estimated values of $\alpha$ are significantly different from the fiducial case. However, note that
the physical observable is not $\alpha$, but $D_{V}/r_{s} = \alpha (D_{V}/r_{s})_{\rm fid}$. Comparing this across
the three cosmologies (second column, Table~\ref{tab:dr7_basic}), we find it insensitive to the choice of 
cosmology. 

\begin{figure}
  \includegraphics[width=3in]{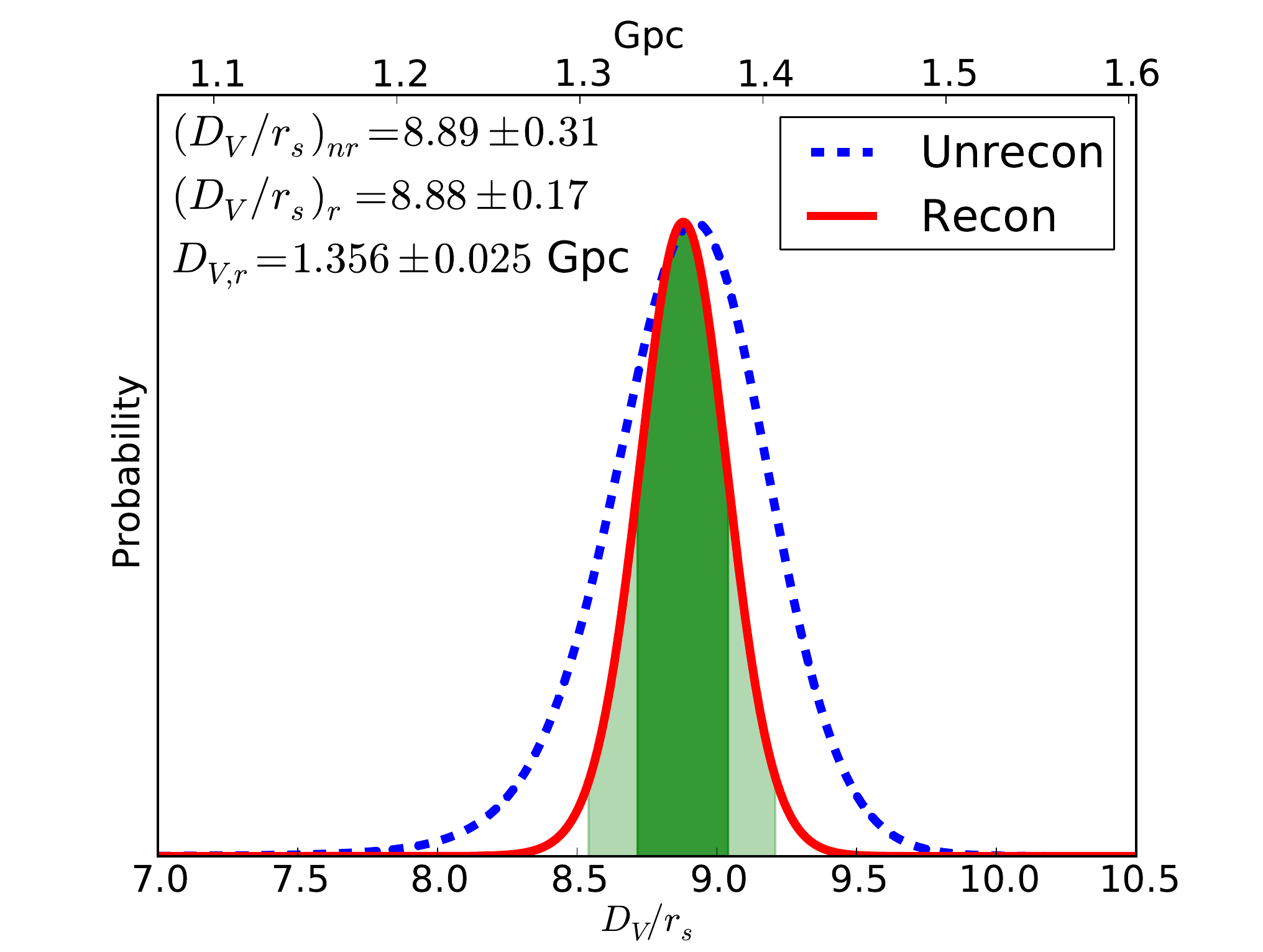}
  \caption{The probability of $\alpha$ before [blue, dashed] and after [red, solid] reconstruction.
  The mean and standard deviations of the two distributions is also listed. 
  The shaded regions show the 1$\sigma$ [dark shaded] and 2$\sigma$ [light shaded] regions of the 
  reconstructed probability distribution.}
  \label{fig:dr7_palpha}
\end{figure}

The distance information from these BAO measurements may be summarized into a probability distribution $p(D_{V}/r_{s})$, 
plotted in Figure~\ref{fig:dr7_palpha} and summarized in the second column of Table~\ref{tab:dr7_basic}. Unlike $\alpha$, 
these measurements no longer make reference to a fiducial cosmology. One may however freely convert between 
$p(\alpha)$ and $p(D_{V}/r_{s})$ by multiplying the latter by $(D_{V}/r_{s})_{\rm fid}$. We use the results in 
Figure~\ref{fig:dr7_palpha} to explore the cosmological consequences of these measurements in Paper III. 
If we assume a perfectly measured sound horizon, these measurements can be converted into a distance measurement in 
Gpc. Using a sound horizon of 152.76 Mpc, we get a distance to $z=0.35$ of $1.356 \pm 0.025$ Gpc. Note that these 
numbers do not have $h^{-1}$ factors in them. Of course, the sound horizon is not perfectly measured and its
uncertainty must be taken into account when fitting for cosmologies. Paper III discusses the methodologies
and results in detail.

\section{Discussion}
\label{sec:discuss}

We present the results of the density field reconstruction on the BAO feature on the SDSS DR7 LRG data. This is 
the {\it first} application of reconstruction on a galaxy redshift survey, resulting in a 1.8 factor reduction in the 
distance error to a $z=0.35$, equivalent to a {\it tripling} of the survey volume. This is the first in a series
of three papers; Paper II describes the fitting of the correlation function, while Paper III explores the 
cosmological implications of these results. 

Our principal results and conclusions are :
\begin{enumerate}
  \item We modify the \cite{2007ApJ...664..675E} reconstruction algorithm to account for the effects of survey boundaries and redshift-space
    distortions and test it on the mock catalogs from the LasDamas suite of simulations. These mock catalogs have been designed to both match the SDSS survey 
    geometry as well as the redshift distribution and clustering properties of the SDSS LRG sample.
  \item We find that reconstruction both sharpens the BAO feature and restores the isotropy of the correlation function on large scales.
    The nonlinear smoothing of the BAO feature decreases from 8.1 to 4.4 Mpc/$h$, in line with theoretical estimates \citep{2009PhRvD..79f3523P}.
  \item We find that reconstruction improves the distance estimates, reducing the median error of the LasDamas simulations from 3.3\% 
    to 2.3\%. Furthermore, reconstruction also significantly reduces the number of outliers as a result of improving the detectability of the 
    BAO feature.
  \item We calibrate the smoothing scale input to the reconstruction algorithm and find that the optimal scale lies between 
    $\sim\!$ 15 to 20 Mpc/$h$; we adopt 15 Mpc/$h$ as our fiducial value. Choosing too small a smoothing scale results in prominent artifacts
    in the correlation function, while too large a scale degrades the efficacy of reconstruction.
  \item We demonstrate the reconstruction is robust to the choices of galaxy bias $b$, the rate of growth of structure $f$ and the 
    fiducial power spectrum used to interpolate missing data and pad the survey edges. 
  \item Applying reconstruction to the SDSS DR7 LRG data, we measure a relative distance to $z=0.35$ of $D_{V}/r_{s}=8.88 \pm 0.17$, compared
    with $8.89 \pm 0.31$ before reconstruction. The two distances are consistent, but reconstruction reduced the error from 3.5\% to 1.9\%,
    a factor of 1.8, equivalent to tripling the survey volume. 
  \item Reconstruction also improves the detectability of a BAO feature (relative to a model with feature erased) from 3.3$\sigma$ to 
    4.2$\sigma$. 
\end{enumerate}

We can compare our results to the results of \citet{2010MNRAS.401.2148P} and \citet{2010ApJ...710.1444K}, who analyzed similar samples 
with different (albeit related) fitting methodologies. 
\citet{2010MNRAS.401.2148P} analyze a combination of the SDSS LRG data as well
as lower redshift data from the SDSS Main galaxy sample. Their primary distance constraints are therefore reported at a lower redshift
from ours: $D_V/r_s(z=0.275) = 7.19\pm0.19$, a $2.7\%$ measurement. Comparing these results with ours requires assuming a model 
to transform to a higher redshift. This comparison is done in detail in Paper III. We can however scale our results to this redshift
(assuming that $\alpha$ does not change significantly with redshift) to obtain  $7.15\pm0.13$.
These results are also consistent with \citet{2010ApJ...710.1444K} who obtain $7.17\pm0.25$.  
We note that our error before reconstruction is larger than the \citet{2010MNRAS.401.2148P} results; this is however expected given 
the somewhat larger volume of that sample. \citet{2010MNRAS.401.2148P} also split their sample into two redshift slices with their higher
redshift slice correspond to our measurements. They obtain $D_V/r_s(z=0.35) = 9.11\pm0.30$, consistent with our results before 
reconstruction.

Our results have important implications for current and future surveys. All of these surveys have assumed some level of reconstruction 
for their projected constraints. This work retires a major risk for these surveys, being the first application to data. Furthermore, 
the degree of reconstruction assumed (a reduction of the nonlinear smoothing scale by 50\%) is consistent with what this work 
has achieved.

This work has also limited itself to the angle averaged correlation function. One of the promises of the BAO method is the ability to 
measure both the angular diameter distance and the Hubble constant. These measurements are complicated by the loss of isotropy in the 
correlation function due to redshift-space distortions. As this paper has shown, reconstruction has the potential to undo the 
effects of redshift-space distortions and could significantly improve measurements of the anisotropic BAO signal. 
We will explore this in future work.

This paper has demonstrated that reconstruction is feasible on data and that it can significantly improve the distance constraints.
We expect that reconstruction will become a standard method for analyzing the BAO signal from large redshift surveys.

Funding for the Sloan Digital Sky Survey (SDSS) and SDSS-II has been provided by the Alfred P. Sloan Foundation, 
the Participating Institutions, the National Science Foundation, the U.S. Department of Energy, the National Aeronautics
and Space Administration, the Japanese Monbukagakusho,  and the Max Planck Society, and the Higher Education Funding 
Council for England. The SDSS Web site is \texttt{http://www.sdss.org/}.

The SDSS is managed by the Astrophysical Research Consortium (ARC) for the Participating Institutions. 
The Participating 
Institutions are the American Museum of Natural History, Astrophysical Institute Potsdam, 
University of Basel, University of 
Cambridge, Case Western Reserve University, The University of Chicago, Drexel University, 
Fermilab, the Institute for Advanced 
Study, the Japan Participation Group, The Johns Hopkins University, the Joint Institute for 
Nuclear Astrophysics, the Kavli 
Institute for Particle Astrophysics and Cosmology, the Korean Scientist Group, the 
Chinese Academy of Sciences (LAMOST), 
Los Alamos National Laboratory, the Max-Planck-Institute for Astronomy (MPIA), the 
Max-Planck-Institute for Astrophysics (MPA), 
New Mexico State University, Ohio State University, University of Pittsburgh, 
University of Portsmouth, Princeton University, 
the United States Naval Observatory and the University of Washington.

We thank the LasDamas collaboration for making their galaxy mock
catalogs public.  We thank Cameron McBride for assistance in using the
LasDamas mocks and comments on earlier versions of this work.
We thank Martin White for useful conversations on reconstruction.
NP and AJC are partially supported by NASA grant NNX11AF43G.
DJE, XX, and KM were supported by NSF grant AST-0707725 and NASA grant NNX07AH11G.
This work was supported in part by the facilities and staff of the Yale University 
Faculty of Arts and Sciences High Performance Computing Center.

\bibliographystyle{mn2elong}
\bibliography{recon,petsc}

\end{document}